\documentclass[prd,twocolumn,showpacs,nofootinbib]{revtex4}
\usepackage{natbib,amsmath,graphicx,bm}

\def\fnl{f_\mathrm{NL}}

\def\mnras{Mon. Not. R. Astron. Soc.}
\def\apjl{Astrophys. J. Lett.}
\def\apjs{Astrophys. J. Suppl.}
\def\physrep{Phys. Rep.}
\def\aj{Astron. J.}
\begin{document}
\title{%
Galaxy-CMB and galaxy-galaxy lensing on large scales: sensitivity to primordial non-Gaussianity
}%
\author{Donghui Jeong}
\email{djeong@astro.as.utexas.edu}
\affiliation{Texas Cosmology Center and Department of Astronomy,
University of Texas at Austin, 1
University Station, C1400, Austin, TX 78712}
\author{Eiichiro Komatsu}
\affiliation{Texas Cosmology Center and Department of Astronomy,
University of Texas at Austin, 1
University Station, C1400, Austin, TX 78712}
\author{Bhuvnesh Jain}
\affiliation{Particle Cosmology Center and Department of Physics and
Astronomy, University of Pennsylvania, Philadelphia, PA 19104}
\begin{abstract}
A convincing detection of primordial non-Gaussianity in the local form
 of the bispectrum, whose amplitude is given by the $\fnl$ parameter,
 offers a powerful test of inflation. In this paper we calculate the
 modification of  two-point cross-correlation statistics of weak lensing
 - galaxy-galaxy lensing and galaxy-Cosmic Microwave Background (CMB)
 cross-correlation - due to $\fnl$. We derive and calculate the
 covariance matrix of galaxy-galaxy lensing including cosmic variance
 terms.  We focus on large scales ($l<100$) for which the shape noise of
 the shear measurement becomes irrelevant and cosmic variance dominates
 the error budget. For a modest degree of non-Gaussianity, $\fnl=\pm
 50$, modifications of the galaxy-galaxy lensing signal at the 10\%
 level are seen on scales $R\sim 300$~Mpc, and grow rapidly toward
 larger scales as $\propto R^2$.  We also see a clear signature of the
 baryonic acoustic oscillation feature in the matter power spectrum at
 $\sim 150$~Mpc, which can be measured by next-generation lensing
 experiments. In addition we can probe the local-form primordial
 non-Gaussianity in the  galaxy-CMB lensing signal by correlating the
 lensing potential reconstructed from CMB with high-$z$ galaxies. For
 example, for $\fnl=\pm 50$, we find that the galaxy-CMB lensing cross
 power spectrum is modified by $\sim 10$\% at $l\sim 40$, and by a
 factor of two at $l\sim 10$, for a population of galaxies at $z=2$ with
 a bias of 2. The effect is greater for more highly biased populations
 at larger $z$; thus, high-$z$ galaxy surveys cross-correlated with CMB
 offer a yet another probe of primordial non-Gaussianity.
\end{abstract}
\maketitle
\section{Introduction}
Why study non-Gaussianity? For many years it was recognized that the
simple inflationary models based upon a single slowly-rolling scalar
field would predict nearly Gaussian primordial fluctuations. In particular,
when we parametrize the magnitude of non-Gaussianity in the
primordial curvature perturbations $\zeta$, which gives the observed
temperature anisotropy in the Cosmic Microwave Background (CMB) in the
Sachs--Wolfe limit as $\Delta T/T=-\zeta/5$, using the so-called
non-linear parameter $\fnl$ \cite{komatsu/spergel:2001} as $\zeta({\mathbf
x})=\zeta_L({\mathbf x})+(3\fnl/5)\zeta_L^2({\mathbf x})$, then the
bispectrum of $\zeta$ is given by\footnote{Definition of the bispectrum in terms of
Fourier coefficients of $\zeta$ is $\langle\zeta_{{\mathbf
k}_1}\zeta_{{\mathbf k}_2}\zeta_{{\mathbf
k}_3}\rangle=(2\pi)^3\delta({\mathbf k}_1+{\mathbf k}_2+{\mathbf
k}_3)B_\zeta(k_1,k_2,k_3)$. Throughout this paper we shall order $k_i$ such
that $k_3\le k_2\le k_1$.}
$B_\zeta(k_1,k_2,k_3)=(6\fnl/5)\left[P_\zeta(k_1)P_\zeta(k_2)+(\mbox{2
			    cyclic terms})\right]$, where
			    $P_\zeta(k)\propto k^{n_s-4}$ is the power
			    spectrum of $\zeta$ and $n_s$ is the tilt of
			    the power spectrum, constrained as
			    $n_s=0.960\pm 0.013$ by the WMAP 5-year data
			    \cite{komatsu/etal:2009}.
This form of the bispectrum has the maximum signal in the so-called squeezed
triangle for which $k_3\ll k_2\approx k_1$ \cite{babich/creminelli/zaldarriaga:2004}. In this limit we obtain
\begin{equation}\label{eq:1}
 B_\zeta(k_1,k_1,k_3\to 0)=\frac{12}5\fnl P_\zeta(k_1)P_\zeta(k_3).
\end{equation}
The earlier calculations showed that $\fnl$ from single-field slow-roll
inflation would be of order the slow-roll parameter, $\epsilon\sim
10^{-2}$
\cite{salopek/bond:1990,falk/rangarajan/srendnicki:1993,gangui/etal:1994}. However,
it is not until recent that it is finally realized that
the coefficient of $P_\zeta(k_1)P_\zeta(k_3)$ from the simplest
single-field slow-roll inflation with the canonical kinetic term
in the squeezed limit is
given precisely by \cite{maldacena:2003,acquaviva/etal:2003}
\begin{equation}
 B_\zeta(k_1,k_1,k_3\to 0)=(1-n_s) P_\zeta(k_1)P_\zeta(k_3).
\label{eq:singleprediction}
\end{equation}
Comparing this result with the form predicted by the $\fnl$ model,
one obtains $\fnl=(5/12)(1-n_s)$.

Perhaps, the most important theoretical discovery regarding primordial
non-Gaussianity from inflation over the last few years is that,
not only models with the canonical kinetic term, but
{\it all}
single-inflation models predict the
bispectrum in the squeezed limit  given by
Eq.~(\ref{eq:singleprediction}), regardless of the form of potential,
kinetic term,
slow-roll, or initial vacuum state
\cite{creminelli/zaldarriaga:2004,seery/lidsey:2005,chen/etal:2007,cheung/etal:2008}.
Therefore, the prediction from all single-field inflation models is
$\fnl=(5/12)(1-n_s)= 0.017$ for $n_s=0.96$.
 A convincing detection of
$\fnl$ well above this level is a breakthrough in our understanding of
the physics of very early universe \cite{bartolo/etal:2004,komatsu/etal:prepb}.
The current limit from the WMAP
5-year data is $\fnl=38\pm 21$ (68\%~CL)
\cite{smith/senatore/zaldarriaga:prep}.

There are many ways of measuring $\fnl$. The most popular method has
been the bispectrum of CMB \cite{verde/etal:2000,wang/kamionkowski:2000,komatsu/spergel:2001,komatsu/etal:2002,komatsu/etal:2003} (also see \cite{komatsu:prep} for a
pedagogical review). The
other methods include the trispectrum of CMB
\cite{okamoto/hu:2002,kogo/komatsu:2006}, the bispectrum of galaxies
\cite{scoccimarro/etal:2004,sefusatti/komatsu:2007,jeong/komatsu:2009b,sefusatti:prep}, and
the abundance of galaxies and clusters of galaxies
\cite{lucchin/matarrese:1988,matarrese/verde/jimenez:2000,sefusatti/etal:2007,loverde/etal:2008}.

Recently, analytical
\citep{dalal/etal:2008,matarrese/verde:2008,slosar/etal:2008,afshordi/tolley:2008,taruya/etal:2008}
and numerical
\citep{dalal/etal:2008,desjacques/seljak/iliev:prep,pillepich/porciani/hahn:prep,grossi/etal:prep} studies
of the effects of primordial non-Gaussianity on the power
spectrum of dark matter halos, $P_h(k)$, have revealed an unexpected
signature of primordial non-Gaussianity in the form of a {\it scale-dependent
galaxy bias}, i.e., $P_h(k)=b_1^2P_m(k)\to [b_1+\Delta b(k)]^2P_m(k)$,
where $P_m(k)$ is the power spectrum of matter density fluctuations, and
\begin{equation}
 \Delta b(k) = \frac{3(b_1-1)\fnl\Omega_mH_0^2\delta_c}{D(z)k^2T(k)}.
\label{eq:bk}
\end{equation}
Here, $D(z)$ and $T(k)$ are the growth rate and the transfer function
for linear matter density fluctuations, respectively,
and $\delta_c=1.68$ is the threshold linear density contrast for a
spherical collapse of an overdensity region.
The $k^2$ factor in the denominator of $\Delta b(k)$ shows that this
effect is important only on very large scales.
Highly biased tracers are more sensitive to $f_{NL}$.

\section{Halo-mass correlation from galaxy-galaxy lensing}
\label{sec:gglens}
\subsection{Formula}

The scale-dependent bias was theoretically discovered when the authors
of \cite{dalal/etal:2008}
studied the form of the {\it cross-correlation} power spectrum between
the dark matter halos and the underlying matter density fluctuations,
$P_{hm}(k)=[b_1+\Delta b(k)]P_m(k)$.
We can observe $P_{hm}(k)$ by cross-correlating the locations of
galaxies or clusters of galaxies with the matter density fluctuations
traced by the weak gravitational lensing (see
\cite{bartelmann/schneider:2001} for a review).

\begin{figure}[t]
\includegraphics[width=85mm]{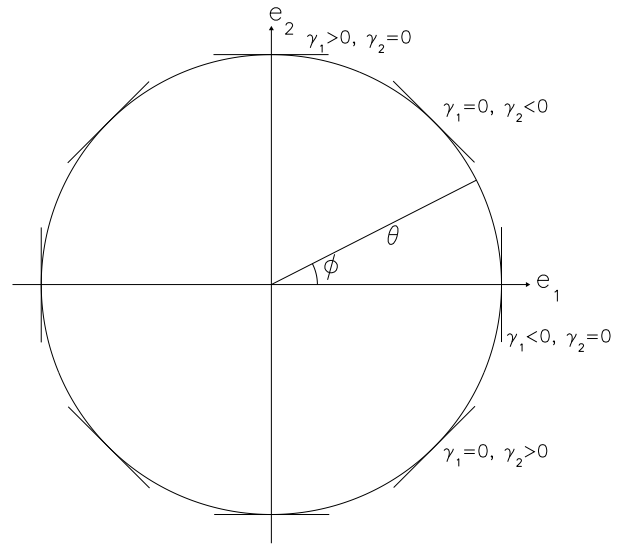}
\caption{Coordinate system and $\gamma_1$ and $\gamma_2$. The shear along ${\bm
 e}_1$ has $\gamma_1>0$ and $\gamma_2=0$, whereas the shear along ${\bm
 e}_2$ has $\gamma_1<0$ and $\gamma_2=0$. The shear along ${\bm
 e}_1+{\bm e}_2$ has $\gamma_1=0$ and $\gamma_2>0$, whereas The shear
 along ${\bm e}_1-{\bm e}_2$ has $\gamma_1=0$ and $\gamma_2<0$.}
\label{fig:gamma}
\end{figure}

One efficient way of measuring $P_{hm}(k)$ is to use
the so-called galaxy-galaxy lensing technique
\cite{tyson/etal:1984,brainerd/brandford/smail:1996,mckay/etal:prep,guzik/seljak:2002,sheldon/etal:2004,mandelbaum/etal:2006,mandelbaum/etal:2006b}:
choose one lens galaxy at a redshift $z_L$, and
measure the mean of {\it tangential}
shears in images of lensed (source or background) galaxies
around the chosen central lensing galaxy as a function of radii from
that central galaxy. Finally, average those mean tangential shears over all
lensing galaxies at
the same redshift, $z_L$.

We begin with the definition of the
tangential shear, $\gamma_t$, on the flat sky\footnote{For an all-sky
analysis, this
relation needs to be replaced with the exact relation using the spin-2
harmonics \cite{stebbins:prep}.}
\begin{equation}
\label{eq:gammat}
 \gamma_t(\bm{\theta})=-\gamma_1(\bm{\theta})\cos(2\phi)-
\gamma_2(\bm{\theta})\sin(2\phi),
\end{equation}
where ${\bm\theta}=(\theta\cos\phi,\theta\sin\phi)$, and
$\gamma_1$ and $\gamma_2$ are components of the  shear field.
\footnote{
As the shear has two independent
components, we are ignoring another linear combination of $\gamma_1$ and
$\gamma_2$
by only focusing on the tangential shear. In particular, on large scales
there is information in the other component of the shear, and thus the
full analysis including both shear components (not just tangential one)
yields a modest (smaller than a factor of $\sqrt{2}$) improvement in the
signal-to-noise ratio. Moreover, using magnification (in addition to
shears), which is proportional to the convergence field $\kappa$, can
also yield a modest improvement.
}
The coordinate system and the meaning of $\gamma_1$ and $\gamma_2$ are
explained in
Fig.~\ref{fig:gamma}. For purely tangential shears shown in
Fig.~\ref{fig:gamma}, $\gamma_t$ is always
positive. This property allows us to average $\gamma_t$ over
the ring around the origin to estimate the mean tangential shear,
$\overline{\gamma}_t$:
\begin{equation}
 \overline{\gamma}_t(\theta) \equiv \int_0^{2\pi}\frac{d\phi}{2\pi}
\gamma_t(\theta,\phi).
\label{eq:mean}
\end{equation}
On the flat sky, $\gamma_1$ and $\gamma_2$ are related to the projected mass
density fluctuation in Fourier space, $\kappa(\mathbf{l})$, as
\begin{eqnarray}
\label{eq:gamma1}
 \gamma_1(\bm{\theta}) &=& \int
  \frac{d^2\mathbf{l}}{(2\pi)^2}\kappa(\mathbf{l})\cos(2\varphi)e^{i\mathbf{l}\cdot\bm{\theta}},\\
\label{eq:gamma2}
 \gamma_2(\bm{\theta}) &=& \int
  \frac{d^2\mathbf{l}}{(2\pi)^2}\kappa(\mathbf{l})\sin(2\varphi)e^{i\mathbf{l}\cdot\bm{\theta}},
\end{eqnarray}
where $\varphi$ is the angle between $\mathbf{l}$ and $\bm{e}_1$,
i.e., $\mathbf{l}=(l\cos\varphi,l\sin\varphi)$.
Using Eqs.~(\ref{eq:gamma1}) and (\ref{eq:gamma2}) in
Eq.~(\ref{eq:gammat}), we write the tangential shear in terms of
$\kappa(\mathbf{l})$ as
\begin{equation}
 \gamma_t(\bm{\theta}) = -\int
  \frac{d^2\mathbf{l}}{(2\pi)^2}\kappa(\mathbf{l})\cos[2(\phi-\varphi)]e^{il\theta\cos(\phi-\varphi)}.
\end{equation}
The mean tangential shear (Eq.~(\ref{eq:mean})) is then given by
\begin{eqnarray}
 \overline{\gamma}_t(\theta) &=&
-\int
  \frac{d^2\mathbf{l}}{(2\pi)^2}\kappa(\mathbf{l})
\int_0^{2\pi}\frac{d\phi}{2\pi}
\cos[2(\phi-\varphi)]e^{il\theta\cos(\phi-\varphi)}\nonumber\\
&=&\int
  \frac{d^2\mathbf{l}}{(2\pi)^2}\kappa(\mathbf{l})
J_2(l\theta).
\label{eq:mean2}
\end{eqnarray}
Here, we have used the identity
\begin{equation}
 J_m(x) = \int_{\alpha}^{2\pi+\alpha}\frac{d\psi}{2\pi}e^{i(m\psi-x\sin\psi)},
\end{equation}
with $m=2$, $\psi=\phi-\varphi-\pi/2$, $\alpha=\varphi+\pi/2$,
and $\int_0^{2\pi}d\psi\sin(2\psi)e^{ix\cos\psi}=0$.

The ensemble average of the mean tangential shear vanishes,
i.e., $\langle \overline{\gamma}_t\rangle=0$, as $\langle\kappa\rangle=0$.
This simply means that the average of the mean tangential shears, measured
with respect to random points on the sky, vanishes.
We obtain non-zero values when we
average the mean tangential shears measured with respect to the locations of
halos (galaxies or clusters of galaxies).
This quantity, called the galaxy-galaxy lensing or cluster-galaxy
lensing, can be used to measure the halo-mass cross correlation.

While clusters of galaxies may be identified directly with dark matter halos
of a given mass, how are galaxies related to halos? Some galaxies (``field
galaxies'') may also be identified directly with dark matter halos;
however, galaxies residing within groups or clusters of galaxies should
be identified with subhalos moving in a bigger dark matter halo. For
such subhalos our argument given below may not be immediately
used. However, it is observationally feasible to identify the central galaxies
in groups or clusters of galaxies and measure the mean tangential shear
around them. A number of studies of Luminous Red Galaxies (LRGs)
extracted from the the Sloan Digital Sky Survey (SDSS) have shown that
these are typical central
galaxies in galaxy groups
\cite{mandelbaum/etal:2006b,sheldon/etal:prepb,johnston/etal:prep}. Scalings
such as the mass-luminosity scaling imply that
LRGs provide a useful proxy for the halos within which they reside. We
will assume in this study that such tracers will enable the halo-shear
cross-correlation to be measured.  There are some caveats such as
bimodal mass distributions in galaxy groups \cite{johnston/etal:prep} and
the extrapolation to higher redshift, but we will leave a detailed
exploration to real galaxy tracers for later work.

The ensemble average of the mean tangential shears relative to the
locations of halos at a given redshift $z_L$, denoted as
$\langle\overline{\gamma}_t^h\rangle(\theta,z_L)$,
 is related to the angular cross-correlation
power spectrum of halos and $\kappa$, $C_l^{h\kappa}$, as \cite{hu/jain:2004}
\begin{equation}
\label{eq:gammath}
 \langle\overline{\gamma}_t^h\rangle(\theta,z_L)
= \int \frac{ldl}{2\pi}C_l^{h\kappa}(z_L)J_2\left(l\theta\right).
\end{equation}
We give the derivation of this result in Appendix~\ref{app:meanshear}.

With the lens redshift $z_L$ known (from spectroscopic observations), we
can calculate the comoving radius,
$R$, corresponding to the angular separation on the sky, $\theta$, as
$R=\theta d_A(0;z_L)$ where $d_A(0;z_L)$ is the {\it comoving} angular
diameter distance from $z=0$ to $z=z_L$.
Using Limber's approximation~\citep{limber:1954,kaiser:1992}
on the flat sky relating $C_l^{h\kappa}$
to $P_{hm}(k)$,\footnote{
As we are dealing with correlations on very
large angular scales, one may worry about the validity of Limber's
approximation. In Appendix~\ref{app:limber} we give a detailed study of
the validity and limitation of Limber's approximation for the
galaxy-galaxy lensing.
}
we can write Eq.~(\ref{eq:gammath}) as \cite{hu/jain:2004}
\begin{equation}
\langle\overline{\gamma}_t^h\rangle(R,z_L)
=
\frac{\rho_0}{\Sigma_c(z_L)}
\int \frac{kdk}{2\pi} P_{hm}(k,z_L)J_2(kR).
\label{eq:gammaT}
\end{equation}
Here,
$\rho_0$ is the mean comoving mass density of the universe, and
$\Sigma_c(z_L)$ is the so-called critical surface density:
\begin{equation}
 \Sigma_c^{-1}(z_L)=\frac{4\pi G}{c^2}(1+z_L)d_A(0;z_L)\int_{z_L}^\infty
  dz_S~p(z_S)\frac{d_A(z_L;z_S)}{d_A(0;z_S)},
\label{eq:sigmac0}
\end{equation}
where
$p(z_S)$ is the redshift distribution of sources normalized to unity,
$\int dzp(z)=1$,
and $d_A(0;z)$ and $d_A(z;z_S)$ are the comoving angular diameter
distances out to $z$ and between $z$ and $z_S$, respectively.
The numerical value of $4 \pi G/c^2$ is
$6.01\times10^{-19}~\mathrm{Mpc/M_\odot}$, and $4\pi G\rho_0/c^2$ is
 $1.67\times 10^{-7} (\Omega_m h^2)~{\rm Mpc}^{-2}$.

\begin{figure}
\begin{center}
\includegraphics[width=9cm]{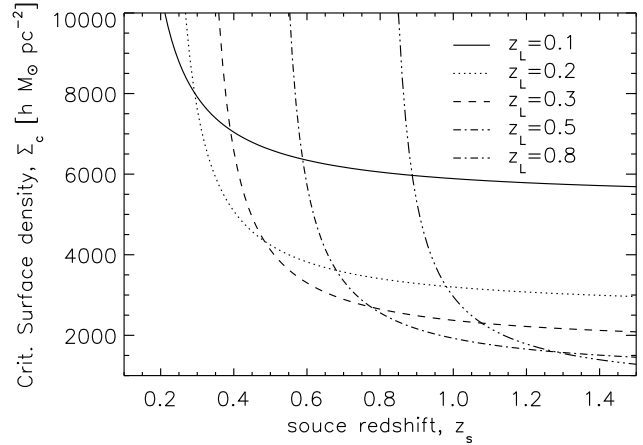}
\end{center}
\caption{
\label{fig:sigma_c}
Critical surface density, $\Sigma_c(z_L;z_S)$, as a function of the
 source redshift, $z_S$, for various lens redshifts that
roughly correspond  to the
 Two Degree Field Galaxy Redshift Survey (2dFGRS; $z_L=0.1$, solid),
 the main sample of the Sloan Digital Sky Survey
 (SDSS; $z_L=0.2$, dotted), the Luminous
 Red Galaxies (LRGs) of SDSS ($z_L=0.3$, dashed), and the Large Synoptic Survey
 Telescope  (LSST; $z_L=0.5$ and $0.8$, dot-dashed and
 triple-dot-dashed, respectively).
}
\end{figure}

Eq.~(\ref{eq:gammaT}) is often written as
\begin{equation}
\langle\overline{\gamma}_t^h\rangle(R,z_L) =
  \frac{\Delta\Sigma(R,z_L)}{\Sigma_c(z_L)}.
\end{equation}
To simplify the analysis, let us define the ``effective source
redshift'' of a given survey from the following equation:
\begin{equation}\label{eq:z_s,eff}
 \frac{d_A(z_L;z_{S,\rm eff})}{d_A(0;z_{S,\rm eff})}\equiv \int_{z_L}^\infty dz_S~p(z_S)\frac{d_A(z_L;z_S)}{d_A(0;z_S)}.
\end{equation}
Henceforth we shall use $z_S$ to denote $z_{S,\rm eff}$, and write
\begin{equation}
 \Sigma_c^{-1}(z_L;z_S)=\frac{4\pi
  G}{c^2}(1+z_L)d_A(0;z_L)\frac{d_A(z_L;z_S)}{d_A(0;z_S)}.
\label{eq:sigmac}
\end{equation}
Fig.~\ref{fig:sigma_c} shows $\Sigma_c$ for $z_L=0.1$ (2dFGRS, Two
Degree Field Galaxy Redshift Survey), 0.2
(SDSS main), 0.3
(SDSS LRG), and 0.5 and 0.8 (both LSST, Large Synoptic Survey Telescope). The smaller $\Sigma_c$ is, the
larger the observed mean tangential shear is.

\subsection{Results}

\begin{figure}
\begin{center}
\includegraphics[width=9cm]{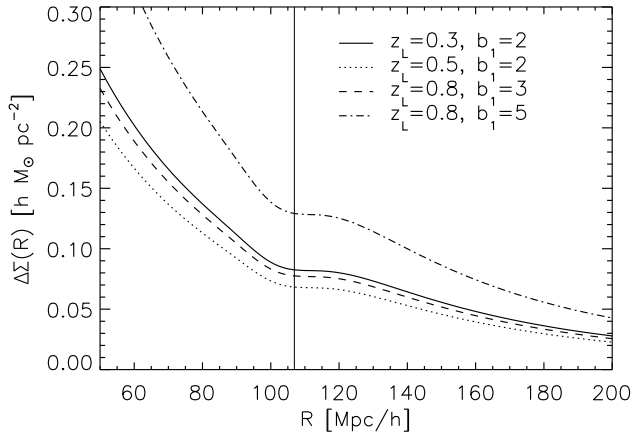}
\end{center}
\caption{
\label{fig:bao}
The baryonic feature in the matter power spectrum, as seen in the
 galaxy-galaxy lensing,
$\Delta\Sigma(R)$, for several populations of lens galaxies with
$b_1=2$ at $z_L=0.3$ (similar to SDSS LRGs, solid),
$b_1=2$ at $z_L=0.5$ (higher-$z$ LRGs, dotted),
$b_1=2$ at $z_L=0.8$ (galaxies that can be observed by LSST, dashed),
and $b_1=5$ at $z_L=0.8$ (clusters of galaxies that can be observed by
 LSST, dot-dashed).
The vertical line shows the location of the baryonic feature, $R_{\rm
 BAO}=106.9~h^{-1}~{
\rm Mpc}$, calculated
 from the ``WMAP+BAO+SN ML'' parameters in
 Table 1 of
\cite{komatsu/etal:2009}. Note that we have used the linear matter
 power spectrum and the Gaussian initial condition ($\fnl=0$) for this calculation.
}
\end{figure}
\begin{figure}
\begin{center}
\includegraphics[width=8cm]{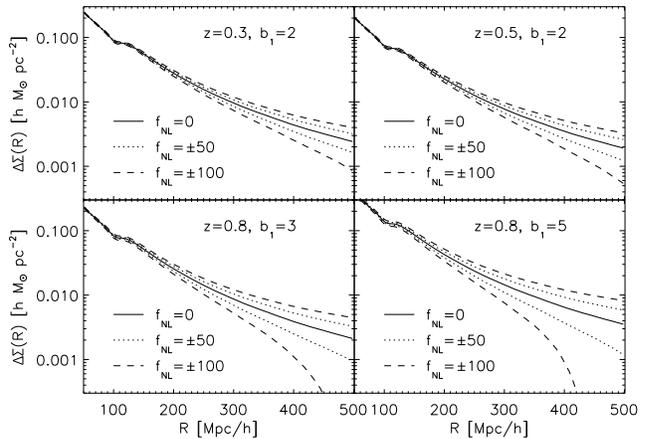}
\end{center}
\caption{
\label{fig:fnl_gglens_signal}
Imprints of the local-type primordial non-Gaussianity in the
 galaxy-galaxy lensing, $\Delta\Sigma(R)$, for the same populations of
 lens galaxies as in Fig.~\ref{fig:bao}.
 The solid, dashed, and dotted lines show $\fnl=0$, $\pm 50$, and $\pm
 100$, respectively.
}
\end{figure}
\begin{figure}
\begin{center}
\includegraphics[width=8cm]{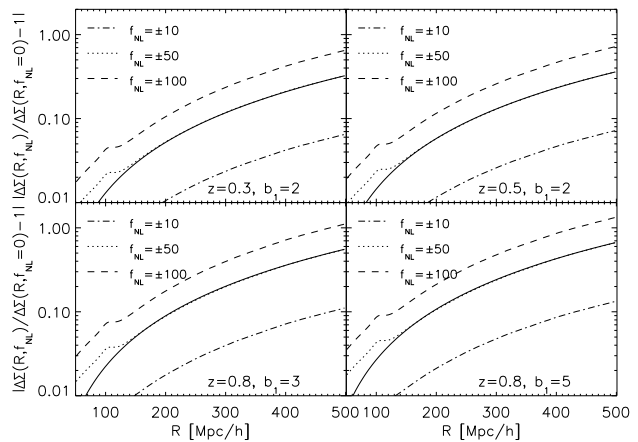}
\end{center}
\caption{
\label{fig:fnl_gglens_diff}
Fractional differences between $\Delta\Sigma(R)$ from non-Gaussian
 initial conditions and the Gaussian initial condition,
 $|\Delta\Sigma(R;\fnl)/\Delta\Sigma(R;\fnl=0)-1|$, calculated from the
 curves shown in Fig.~\ref{fig:fnl_gglens_signal}.
  The  dot-dashed, dashed, and dotted lines show $\fnl=\pm 10$, $\pm
 50$, and $\pm
 100$, respectively, while the thin solid line shows $\propto R^2$ with an
 arbitrary normalization.
}
\end{figure}

We can now calculate the observable, $\Delta\Sigma(R,z_L)$, for various
values of $\fnl$. We use
\begin{eqnarray}
\nonumber
& &\Delta\Sigma(R,z_L)\\
\nonumber
&=&
\rho_0b_1
\int \frac{kdk}{2\pi}P_{m}(k,z_L)J_2(kR)\\
& &+
\rho_0
\int \frac{kdk}{2\pi}
\Delta b(k,z_L)P_{m}(k,z_L)J_2(kR),
\label{eq:DeltaSigma}
\end{eqnarray}
where the scale-dependent bias, $\Delta b(k,z)$, is given by
Eq.~(\ref{eq:bk}). As we are interested in large scales, i.e.,
$R>10~h^{-1}~{\rm Mpc}$, we shall use the linear matter spectrum for $P_m(k)$.

Fig.~\ref{fig:bao} shows, for the Gaussian initial condition ($\fnl=0$),
$\Delta\Sigma(R,z_L)$ from $R=50$ to $200~{h^{-1}~{\rm Mpc}}$.
We have chosen the bias parameters and lens redshifts to represent the
existing data sets as well as the future ones:
$b_1=2$ at $z_L=0.3$ (similar to
the observed values from SDSS LRGs
\citep{tegmark/etal:2006},
top-left),
$b_1=2$ at $z_L=0.5$ (higher-$z$
LRGs
\citep{schlegel/etal:prep},
top-right),
$b_1=2$ at $z_L=0.8$ (galaxies that can be observed by LSST,
\citep{zhan:2006},
bottom-left),
and $b_1=5$ at $z_L=0.8$ (clusters of galaxies that can be observed by
LSST, bottom-right).
While LSST is an imaging survey, we assume that we can obtain
spectroscopic redshifts of
some ($\sim 10^6$) lens galaxies by follow-up observations.
It is also straightforward to extend our analysis to lenses selected by
photometric redshifts.

At $R\sim 110~h^{-1}~{\rm Mpc}$ we see a clear ``shoulder'' due to the baryonic
feature in the linear matter power spectrum (often called Baryon
Acoustic Oscillations; BAO). The sound horizon at the drag epoch (which
is more relevant to the matter power spectrum than the photon decoupling
epoch for the CMB power spectrum) calculated from the cosmological model
that we use,  the ``WMAP+BAO+SN ML'' parameters in
 Table 1 of \cite{komatsu/etal:2009}, is 106.9~$h^{-1}$~Mpc, as shown as
 the vertical line in this figure.
The magnitude of $\Delta\Sigma$ on this
scale is $\sim 0.1~h~M_\odot~{\rm pc}^{-2}$.
Assuming a range of $\Sigma_c$ from future surveys,
$\Sigma_c\sim 1000-4000~h~M_\odot~{\rm pc}^{-2}$ (see Fig.~\ref{fig:sigma_c}), this value
corresponds to the mean tangential shear of order $2.5\times 10^{-5}$ to
$10^{-4}$. Is this observable?

For comparison, Sheldon et al.~\cite{sheldon/etal:prepb} measured
$\Delta\Sigma(R)\sim 0.5~h~M_\odot~{\rm pc}^{-2}$ at $R\sim
30~h^{-1}~{\rm Mpc}$ from clusters of galaxies in the SDSS
main sample. The mean lens redshift for these data is $z_L\sim 0.2$,
which would give $\Sigma_c\sim 5000~h~M_\odot~{\rm pc}^{-2}$
(see Fig.~\ref{fig:sigma_c} for $z_L=0.2$ and $z_S\sim 0.4$); thus, the magnitude of
 the mean tangential shear that they were able to measure is of order
$10^{-4}$, which is only $\sim 1$ to 4 times larger than the magnitude of the
signal expected from the BAO. Therefore, detecting the BAO
signature in $\Delta\Sigma(R)$ should be quite feasible with the future
observations. We shall give a more quantitative discussion on the detectability of BAO from the galaxy-galaxy
lensing effect  in Sec~\ref{sec:gglens_detectability}.

How about $\fnl$? As expected, the effect of $\fnl$ is enhanced on very
large scales, i.e., hundreds of Mpc (see Fig.~\ref{fig:fnl_gglens_signal}).
For $\fnl=\pm 50$,
$\Delta\Sigma(R)$ is modified by 10--20\% at $R\sim
300~h^{-1}~{\rm Mpc}$ (depending on $b_1$ and $z_L$; see
Fig.~\ref{fig:fnl_gglens_diff}). The modification grows rapidly toward
larger scales, in proportion to $R^2$. On such a large
scale ($R\sim
300~h^{-1}~{\rm Mpc}$ ), the galaxy-galaxy lensing signal is on the
order of $\Delta\Sigma\sim 0.01~h~M_\odot~{\rm pc}^{-2}$, and thus we need
to measure the mean tangential shear down to the level of
$\overline{\gamma}_t^h\sim 2.5\times 10^{-6}$ to $10^{-5}$, i.e.,
10--40 times smaller than the level of sensitivity achieved by the
current observations.
Can we observe such a small shear?

\subsection{Covariance matrix of the mean tangential shear}

In order to study the feasibility of measuring the tangential shear of
order $10^{-6}$, we compute the covariance matrix of the mean
tangential shears averaged over $N_L$ lens galaxies.
As derived in Appendix~\ref{app:covariance},
the covariance matrix of the mean tangential shear is
\begin{eqnarray}
 \nonumber
&
&\langle\overline{\gamma}_t^h(\theta)\overline{\gamma}_t^h(\theta')\rangle
-\langle\overline{\gamma}_t^h(\theta)\rangle\langle\overline{\gamma}_t^h(\theta')\rangle
\\
\nonumber
&=&
\frac1{4\pi f_{\rm sky}}
\int \frac{ldl}{2\pi}J_2(l\theta)J_2(l\theta')\\
& &\times
\left[
(C_l^{h\kappa})^2+
\left(
C_l^h+\frac1{n_L}
\right)
\left(
C_l^\kappa+\frac{\sigma_\gamma^2}{n_S}
\right)
\right].
\label{eq:cov}
\end{eqnarray}
This expression includes the cosmic variance, the shot noise of lens halos,
as well as the shape noise $\sigma_\gamma$. As far as we know this
formula has not been derived before.
Note that we have assumed a single source and lens redshift. For
multiple source and lens redshifts, the covariance matrix needs to be
suitably generalized.

Here, $C_l^{h}$ and $C_l^{\kappa}$ are the angular power spectra of the
lens halos (galaxies or cluster of galaxies) and $\kappa$, respectively,
and $n_L$ and $n_S$ are the number densities of the lens halos and the
lensed (source) galaxies, respectively.
These angular power spectra,
$C_l^{h\kappa}$, $C_l^h$, $C_l^{\kappa}$, will be related to
the corresponding three-dimensional power
spectrum, $P(k)$, in Sec~\ref{sec:Clgk_cov}.

In the limit that the cosmic
variance is unimportant, we recover the usual expression used in the
literature:
\begin{eqnarray}
\langle\overline{\gamma}_t^h(\theta)\overline{\gamma}_t^h(\theta')\rangle
-\langle\overline{\gamma}_t^h(\theta)\rangle\langle\overline{\gamma}_t^h(\theta')\rangle
=
\frac{\sigma_\gamma^2}{N_L}\frac{\delta_D(\theta-\theta')}{2\pi\theta n_S},
\end{eqnarray}
where $N_L=4\pi f_{\rm sky}n_L$ is the total number of lens halos available in the data.
In this limit the errors in different radial bins are uncorrelated,
and they are simply given by the shape noise, $\sigma_\gamma$, reduced
by the square-root of the number of source galaxies available within
each radial bin and the total number of lens halos that we can use for
averaging the mean tangential shear. In particular, at each bin with a width
$\Delta\theta$, we find the variance of
\begin{eqnarray}
{\rm Var}[\overline{\gamma}_t^h(\theta)]
=
\frac{\sigma^2_\gamma}{2\pi\theta(\Delta\theta) n_SN_L},
\end{eqnarray}
in the absence of the cosmic variance.

When would the cosmic variance become important?
There is the maximum surface number density of sources,
$n_{S,\rm max}=\sigma_\gamma^2/C_l^\kappa$,
 above which the
shape noise becomes irrelevant.
This gives the maximum number of sources within a given radial bin of
a width $\Delta\theta$ ($\ll \theta$) above which the shape noise
becomes irrelevant:
\begin{equation}
 N_{S,\rm max}=2\pi\theta (\Delta\theta) n_{S,\rm max}
=
(l\theta)^2\left(\frac{\Delta\theta}{\theta}\right)\frac{\sigma_\gamma^2}{l^2C_l^\kappa/(2\pi)}.
\end{equation}
For $l\theta= \pi$ (the usual relation between $l$ and $\theta$) and $\sigma_\gamma\simeq 0.3$ (realistic shape
noise), we find
\begin{equation}
 N_{S,\rm max}\simeq
\left(\frac{\Delta\theta}{\theta}\right)\frac1{l^2C_l^\kappa/(2\pi)}.
\end{equation}
At $l\sim 100$, $l^2C_l^\kappa/(2\pi)\sim 10^{-5}$ \cite{hu/jain:2004};
thus, we do not gain sensitivity any further by having more than, say,
$10^4$ galaxies (for $\Delta\theta/\theta=0.1$) within a single radial bin.

Alternatively, one can define the minimum multipole, $l_{\rm min}$,
below which the cosmic variance term dominates:
\begin{equation}
 l_{\rm min}=\sqrt{\frac{2\pi n_S}{\sigma_\gamma^2}\frac{l^2C_l^\kappa}{2\pi}}.
\end{equation}
For LSST, we expect to have the surface density of sources on the order
of $n_S=30~{\rm arcmin}^{-2}=3.5\times 10^8~{\rm sr}^{-1}$. For
$\sigma_\gamma=0.3$, we find
$l_{\rm min}({\rm LSST})\sim 1.6\times 10^5\sqrt{l^2C_l^\kappa/(2\pi)}$.
At $l\lesssim 10^3$, $l^2C_l^\kappa/(2\pi)\lesssim 10^{-4}$
\cite{hu/jain:2004}; thus, at $l\lesssim 10^3$ the cosmic variance term
dominates.

In the limit that the covariance matrix is dominated by the cosmic
variance terms, we have
\begin{eqnarray}
 \nonumber
&
&\langle\overline{\gamma}_t^h(\theta)\overline{\gamma}_t^h(\theta')\rangle
-\langle\overline{\gamma}_t^h(\theta)\rangle\langle\overline{\gamma}_t^h(\theta')\rangle
\\
\nonumber
&=&
\frac1{4\pi f_{\rm sky}}
\int \frac{ldl}{2\pi}J_2(l\theta)J_2(l\theta')
C_l^h
C_l^\kappa(1+r_l^2),
\end{eqnarray}
where $r_l\equiv C_l^{h\kappa}/\sqrt{C_l^\kappa C_l^h}$ is the
cross-correlation coefficient. The variance at a given radial bin is
\begin{eqnarray}
 \nonumber
&
&
\rm{Var}[\overline{\gamma}_t^h(\theta)]
\\
&=&
\frac1{4\pi f_{\rm sky}}
\int \frac{ldl}{2\pi}[J_2(l\theta)]^2
C_l^h
C_l^\kappa(1+r_l^2).
\end{eqnarray}

\subsection{Detectability of the mean tangential shear}
\label{sec:gglens_detectability}

\begin{figure}
\begin{center}
\includegraphics[width=8cm]{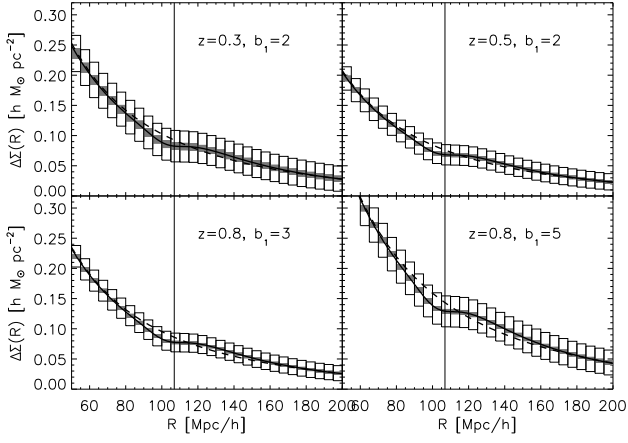}
\end{center}
\caption{
\label{fig:bao_ebar}
Same as Fig.~\ref{fig:bao}, but with the expected 1-$\sigma$
 uncertainties for full-sky lens surveys and a single lens redshift.
 Adjacent bins are highly correlated, with the correlation
 coefficients shown in Fig.~\ref{fig:bao_cov}.
The open (filled) boxes show the binned uncertainties
 with (without) the cosmic variance term due to the cosmic shear field
 included. See Eq.~(\ref{eq:binned_err_tot}) and (\ref{eq:binned_err_shape})
 for the formulae giving open and filled boxes, respectively.
We use the radial bin of size $\Delta R=5~h^{-1}~\mathrm{Mpc}$.
For comparison, we also show $\Delta\Sigma(R)$ computed from the smooth
 power spectrum without the baryonic feature \citep{eisenstein/hu:1998}
 (dashed lines).
Note that the uncertainties are calculated for a single lens redshift
 slice, and thus they will go down as we add more lens redshift slices.
}
\end{figure}
\begin{figure}
\begin{center}
\includegraphics[width=9cm]{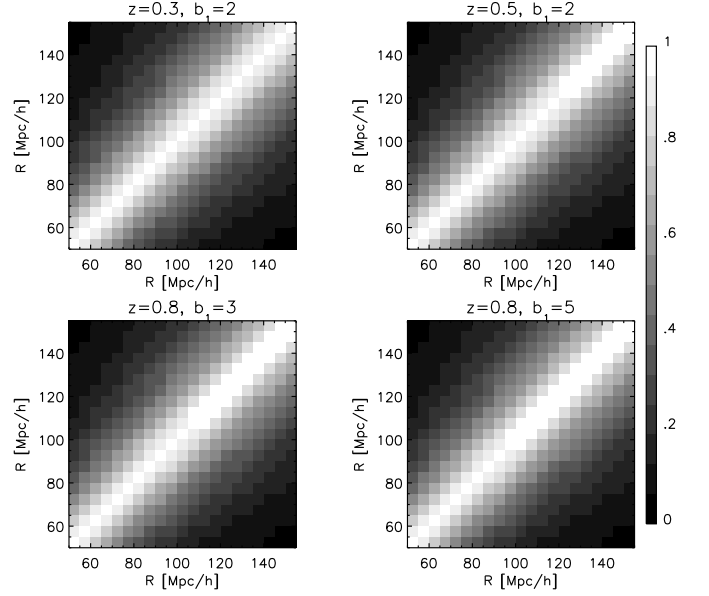}
\end{center}
\caption{
\label{fig:bao_cov}
The cross-correlation-coefficient matrix,
$r_{ij}\equiv C_{ij}/\sqrt{C_{ii}C_{jj}}$, where $C_{ij}$ is the
covariance matrix given in Eq.~(\ref{eq:binned_cov}),
for a radial bin of $\Delta R=5~h^{-1}~\mathrm{Mpc}$. We show $r_{ij}$
for the same populations of lens galaxies as shown in Fig.~\ref{fig:bao} and
\ref{fig:bao_ebar}.
We use the same number of source galaxies and the same shape noise as in
Fig.~\ref{fig:bao_ebar}. The neighboring bins are highly correlated for
$\Delta R < 10~h^{-1}~\mathrm{Mpc}$.
}
\end{figure}
\begin{figure}
\begin{center}
\includegraphics[width=8cm]{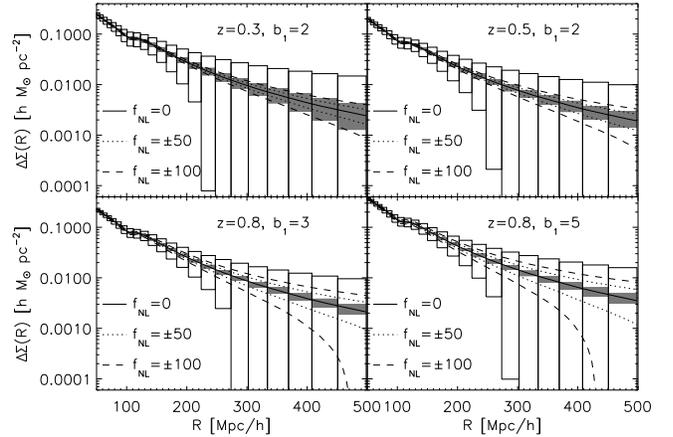}
\end{center}
\caption{
\label{fig:Dsigma_ebar}
Same as Fig.~\ref{fig:fnl_gglens_signal},
but with the expected 1-$\sigma$
 uncertainties for full-sky lens surveys and a single lens redshift.
Adjacent bins are highly correlated.
The open (filled) boxes show the binned uncertainties
 with (without) the cosmic variance term due to the cosmic shear field
 included. See Eq.~(\ref{eq:binned_err_tot}) and (\ref{eq:binned_err_shape})
 for the formulae giving open and filled boxes, respectively.
We use logarithmic bins with  $\Delta R=R/10$.
Note that the uncertainties are calculated for a single lens redshift
 slice, and thus they will go down as we add more lens redshift slices.
}
\end{figure}

In this section, we shall calculate
the expected uncertainties in radially binned
measurements of the mean tangential shear.

The mean tangential shear averaged within the $i$-th bin,
$\langle\hat{\overline{\gamma}}_t^h\rangle(\theta_i)$, i.e.,
the mean tangential shear averaged within an annulus  between
$\theta_{i,\mathrm{min}}$ and $\theta_{i,\mathrm{max}}$, is given by
\begin{eqnarray}
\nonumber
\langle\hat{\overline{\gamma}}_t^h\rangle(\theta_i)
&=&
\frac{2\pi}{A(\theta_i)}
\int_{\theta_{i,\mathrm{min}}}^{\theta_{i,\mathrm{max}}}\theta
d\theta~
\langle\overline{\gamma}_t^h\rangle(\theta)
\\
&\equiv&
\int\frac{ldl}{2\pi}C_l^{h\kappa}\hat{J}_2(l\theta_i),
\end{eqnarray}
where
$A(\theta_i)=
\pi(\theta^2_{i,\mathrm{max}}-\theta^2_{i,\mathrm{min}})$ is the area of
the annulus, and
\begin{equation}
\hat{J}_2(l\theta_i)
=
\frac{2\pi}{A(\theta_i)}
\int_{\theta_{i,\mathrm{min}}}^{\theta_{i,\mathrm{max}}}
\theta d\theta~  J_2(l\theta),
\end{equation}
is the Bessel function averaged within a bin.

Similarly, the covariance matrix of the binned mean tangential shears is
given by
\begin{eqnarray}
 \nonumber
C_{ij}
&
\equiv
&\langle\hat{\overline{\gamma}}_t^h(\theta_i)\hat{\overline{\gamma}}_t^h(\theta_j)\rangle
-\langle\hat{\overline{\gamma}}_t^h(\theta_i)\rangle\langle\hat{\overline{\gamma}}_t^h(\theta_j)\rangle
\\
\nonumber
&=&
\frac1{4\pi f_{\rm sky}}
\int \frac{ldl}{2\pi}\hat{J}_2(l\theta_i)\hat{J}_2(l\theta_j)
\\
& &\times
\left[
(C_l^{h\kappa})^2+
\left(
C_l^h+\frac1{n_L}
\right)
\left(
C_l^\kappa+\frac{\sigma_\gamma^2}{n_S}
\right)
\right].
\label{eq:binned_cov}
\end{eqnarray}
This matrix contains the full information regarding the statistical
 errors of the binned measurements of the mean tangential shear, which
 includes  the cosmic variance errors due to the cosmic shear
 ($C_l^{\kappa}$), clustering of lens galaxies ($C_l^h$) and their
 correlations ($C_l^{h\kappa}$), the finite
 number density of lenses,  and the noise in intrinsic shapes of source
 galaxies.

The variance at a given radial bin is
\begin{eqnarray}
\label{eq:binned_err_tot}
 \nonumber
\mathrm{Var}[\hat{\overline{\gamma}}_t^h(\theta_i)]
&=&
\frac1{4\pi f_{\rm sky}}
\int \frac{ldl}{2\pi}[\hat{J}_2(l\theta_i)]^2
\\
&\times&
\left[
(C_l^{h\kappa})^2+
\left(
C_l^h+\frac1{n_L}
\right)
\left(
C_l^\kappa+\frac{\sigma_\gamma^2}{n_S}
\right)
\right].
\end{eqnarray}
In the analysis of the galaxy-galaxy lensing effects in the literature,
the cosmic variance due to cosmic shear is usually ignored:
\begin{eqnarray}
\nonumber
\left.{\rm Var}[\hat{\overline{\gamma}}_t^h(\theta_i)]\right|_{\kappa=0}
&=&
\frac1{4\pi f_{\rm sky}}
\int \frac{ldl}{2\pi}\left[\hat{J}_2(l\theta_i)\right]^2\\
&\times&
\left[
\left(
C_l^h+\frac1{n_L}
\right)
\frac{\sigma_\gamma^2}{n_S}
\right].
\label{eq:binned_err_shape}
\end{eqnarray}
This is probably a reasonable approximation for the current measurements
at $R\lesssim
30~h^{-1}~{\rm Mpc}$; however, on larger scales which will be probed by
the next-generation lens surveys, the cosmic variance due to cosmic
shear must be included, as we show in Fig.~\ref{fig:bao_ebar}.

For estimating the expected uncertainties, we assume a million lens
galaxies with very narrow (delta-function like) redshift distribution
centered at $z_L$ ($N_L=10^6$) over the full sky, $f_{\rm sky}=1$. We
also assume $\sigma_\gamma=0.3$, and
$n_S=3.5\times 10^{8}~\mathrm{sr}^{-1}$.
As the covariance matrix is dominated by the
 cosmic variance terms, the size of open boxes is insensitive to the
 exact values of $N_L$, $\sigma_\gamma$, or $n_S$.
 (See Sec.~\ref{sec:Clgk_cov}.)
First, we calculate the binned uncertainties in the region close to the
baryonic feature, $R\sim 110~h^{-1}~{\rm Mpc}$.
In Fig.~\ref{fig:bao_ebar},
the open boxes show the full uncertainties including the cosmic variance
due to cosmic shear (Eq.~(\ref{eq:binned_err_tot})), while the
filled boxes show the uncertainties without the cosmic shear term
(Eq.~(\ref{eq:binned_err_shape})). The latter is clearly negligible
compared to the former on large scales, $R\gtrsim 50~h^{-1}~{\rm Mpc}$.

Can we distinguish  $\Delta \Sigma(R)$  with and without the baryonic
feature? Without baryons, we do not see any features in $\Delta
\Sigma(R)$; see dashed lines in Fig.~\ref{fig:bao_ebar} which are
calculated from the
smooth linear power spectrum without the baryonic feature
\citep{eisenstein/hu:1998}. To see if we can detect this feature in
$\Delta \Sigma(R)$,  we estimate the $\chi^2$ difference between
$\Delta \Sigma(R)$  with and without the baryonic
feature:
$$
\Delta \chi^2
\equiv
\sum_{i,j}
(\Delta\Sigma_i - \Delta\Sigma_{i,\mathrm{nw}})
C^{-1}_{ij}
(\Delta\Sigma_j - \Delta\Sigma_{j,\mathrm{nw}}),
$$
where
$\Delta\Sigma_i$ is the mean tangential shear of $i$-th bin,
$\Delta\Sigma_{\mathrm{nw}}$ is $\Delta\Sigma$ without the baryonic
feature, and
$C^{-1}_{ij}$ is the inverse of the binned covariance matrix
(Eq.~\ref{eq:binned_cov}).
Using only a single lens redshift slice, we find
$\Delta\chi^2 = 0.85$ ($z_L=0.3$, $b=2$),
$1.07$ ($z_L=0.5$, $b=2$), $1.32$ ($z_L=0.8$, $b=3$),
and $1.34$ ($z_L=0.8$, $b=5$). For example, if we add up all these
measurements at different slices ($z_L=0.3$, 0.5 and 0.8), significance of detection of the
baryonic feature is $\Delta\chi^2=3.2$, i.e., 93\%~C.L.
As we expect to have many more lens redshift slices from the future lens
surveys, detection and measurement of the baryonic feature in
$\Delta\Sigma$ are quite feasible.
For multiple lens slices the gain in the signal-to-noise ratio will be
approximately $\sqrt{N_{\rm lens}}$; thus, for 10 lens slices the
errors would be a factor of 3 smaller. At best we can expect $\sim 25$
slices, which gives a factor of 5 reduction in errors.

What about $\fnl$? We show the expected 1-$\sigma$ uncertainties for the
mean tangential shears, $\Delta\Sigma(R)$, on larger scales in
Fig.~\ref{fig:Dsigma_ebar}. For this figure we use logarithmic bins with
the radial size of $\Delta R/R=0.1$.
We find that $\Delta\Sigma(R)$ on $R \simeq 250~h^{-1}~\mathrm{Mpc}$ is
detectable, even from a single lens redshift slice. This is remarkable;
however, the
predicted uncertainties are too large for us to distinguish between
$\fnl=0$ and
$\fnl=100$ using a single lens redshift slice. In order to obtain
a tight limit on $\fnl$, we would need to include many lens redshift
slices.

Note that the uncertainty at a given $R$ is larger for  a smaller
lens redshift. This is because a given $R$ corresponds to
a larger angular size for a lower lens redshift, making the cosmic
variance contribution greater.

\section{Harmonic Space Approach}
\label{sec:Clgk}
\subsection{Formula}
\begin{figure}[t]
\begin{center}
\includegraphics[width=9cm]{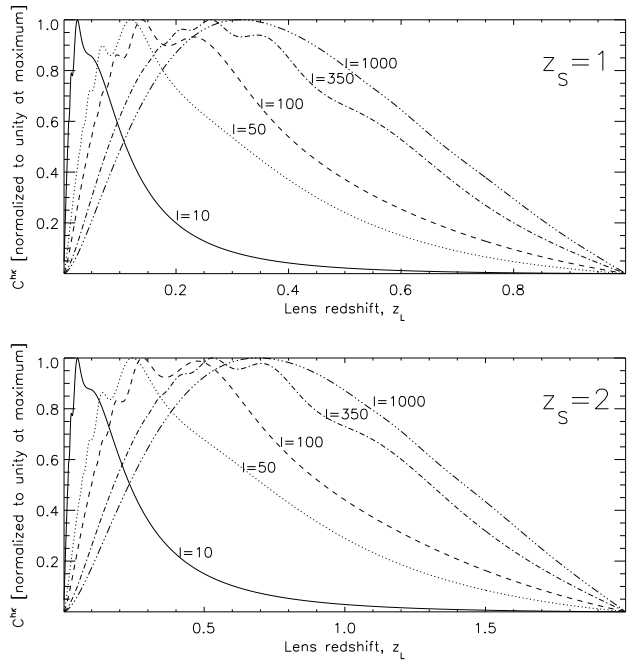}
\end{center}
\caption{
\label{fig:cl_gglens_redshift}
Angular power spectrum of the galaxy-convergence cross correlation,
 $C_l^{h\kappa}$, at various multipoles as a function of the lens
 redshift, $z_L$, for two effective source redshifts, $z_s=1$ (top)
 and $2$ (bottom). We have divided $C_l^{h\kappa}$ by its maximum value.
 The solid, dotted, dashed, dot-dashed, and triple-dot-dashed
 lines show $l=10$, 50, 100, 350, and 1000, respectively.
}
\end{figure}
The mean tangential shear,
$\langle\overline{\gamma}_t^h\rangle$ or $\Delta\Sigma$, is currently
widely used for measuring the halo-shear cross correlation, as this
method is easy to implement and is less sensitive to systematic errors.

In this section, we shall study the effects of $\fnl$ on the equivalent
quantity in harmonic space: the halo-convergence cross power spectrum,
$C_l^{h\kappa}$. The mean tangential shear is related to $C_l^{h\kappa}$
by the 2-dimensional Fourier integral given
in Eq.~(\ref{eq:gammath}).

The convergence field, $\kappa(\mathbf{n})$, is the matter density
fluctuations projected on the sky:
\begin{equation}
\kappa(\mathbf{n})
=
\int_0^{\infty} dz W_\kappa(z)\delta_m[d_A(0;z)\mathbf{n},z],
\end{equation}
where $\delta_m(\mathbf{r},z)\equiv
\rho_m(\mathbf{r},z)/\bar{\rho}_m(z)-1$, and $W_\kappa(z)$ is a lens
kernel which describes the efficiency of lensing for a given redshift
distribution of sources, $p(z_S)$:
\begin{equation}
W_\kappa(z)
=
\frac{\rho_0}{\Sigma_c(z;z_S)H(z)},
\end{equation}
where the critical density, $\Sigma_c$, is defined in
Eq.~(\ref{eq:sigmac}).

Again using Limber's approximation
(whose validity and limitation are studied in
Appendix~\ref{app:limber}), we find the relation between
the angular cross-correlation power spectrum of the convergence field
and the halo density at a given lens redshift $z_L$,
$C_l^{h\kappa}(z_L)$,
and the halo-mass cross-correlation power spectrum at the same redshift,
$P_{hm}(k,z_L)$, as
\begin{eqnarray}
\nonumber
C_l^{h\kappa}(z_L)
&=&
\frac{\rho_0}{\Sigma_c(z_L;z_S)d_A^2(0;z_L)}
P_{hm}\left[k=\frac{l+1/2}{d_A(0;z_L)},z_L\right]
\\
\nonumber
&=& \frac{4\pi G\rho_0}{c^2}
(1+z_L)\frac{d_A(z_L;z_S)}{d_A(0;z_L)d_A(0;z_S)}\\
& &\times P_{hm}\left[k=\frac{l+1/2}{d_A(0;z_L)},z_L\right].
\label{eq:Clhk_limber}
\end{eqnarray}

Fig.~\ref{fig:cl_gglens_redshift} shows $C_l^{h\kappa}(z_L)$ for the
Gaussian density field as a function of lens redshifts, $z_L$.
The convergence fields
at low (high) multipoles are better correlated with low-$z$ (high-$z$)
galaxies. This is due to the shape of the matter power spectrum: on very
large scales (i.e., low $l$), the matter power spectrum is given by the
initial power spectrum, $P_{hm}(k)\propto k$, and thus
we get $1/d_A(0;z_L)$ from $P_{hm}[k=l/d_A(0;z_L)]$.
This gives a larger weight to low-$z$ galaxies.
On smaller scales where
$P_{hm}(k)\propto k^{n_{\rm eff}}$ with $n_{\rm eff}\simeq-3$, we get positive
powers of $d_A(0;z_L)$ from $P_{hm}[k=l/d_A(0;z_L)]$, which gives a
larger weight to high-$z$ galaxies.

\subsection{Result}
\begin{figure}[t]
\begin{center}
\includegraphics[width=8cm]{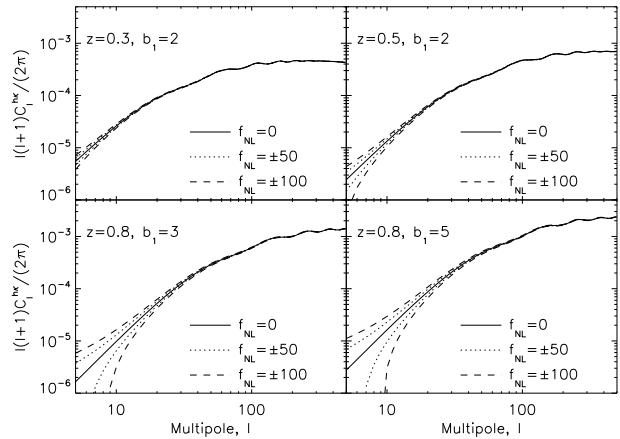}
\end{center}
\caption{
Imprints of the local-type primordial non-Gaussianity in the
 galaxy-convergence cross power spectrum,
 $l(l+1)C_l^{h\kappa}/(2\pi)$, for
for the same populations of
 lens galaxies as in Fig.~\ref{fig:bao}.
 The solid, dashed, and dotted lines show $\fnl=0$, $\pm 50$, and $\pm
 100$, respectively.
\label{fig:cl_gglens_signal}
}
\end{figure}
\begin{figure}
\begin{center}
\includegraphics[width=8cm]{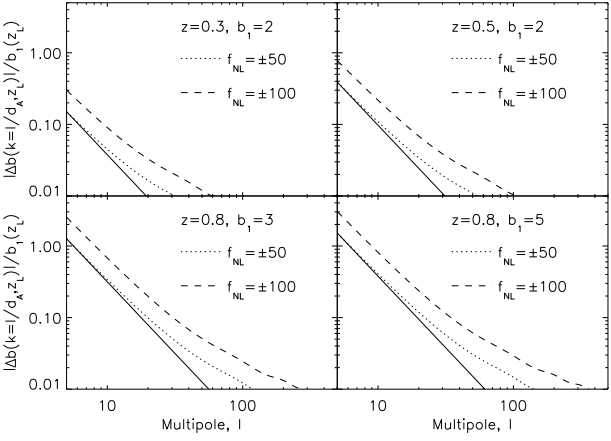}
\end{center}
\caption{
\label{fig:cl_gglens_diff}
Fractional differences between $C_l^{h\kappa}$ from non-Gaussian
 initial conditions and the Gaussian initial condition, calculated from
 the curves shown in Fig.~\ref{fig:cl_gglens_signal}.
These differences are equal
 to  $|\Delta b(l=k/d_A,z_L)|/b_1(z_L)$.
  The  dashed and dotted lines show $\fnl=\pm 50$ and $\pm
 100$, respectively, while the thin solid lines show $l^{-2}$ with an
 arbitrary normalization.
}
\end{figure}
We can now calculate $C_l^{h\kappa}$ for various values of $\fnl$. We
use
\begin{eqnarray}
\nonumber
 C_l^{h\kappa}(z_L)
&=&
\frac{4\pi G\rho_0}{c^2}(1+z_L)\frac{d_A(z_L;z_S)}{d_A(0;z_L)d_A(0;z_S)}\\
\nonumber
& &\times
\left[b_1(z_L)+\Delta b\left(k=\frac{l+1/2}{d_A(0;z_L)},z_L\right)\right]\\
& &\times
P_{m}\left[k=\frac{l+1/2}{d_A(0;z_L)},z_L\right],
\end{eqnarray}
where the scale-dependent bias, $\Delta b(k,z)$, is given by
Eq.~(\ref{eq:bk}).

Figure~\ref{fig:cl_gglens_signal} shows $C_l^{h\kappa}(z_L)$ for $\fnl=\pm
50$ and $\pm 100$ for populations of galaxies that we have
considered in the previous sections. For each lens redshift, we
calculate the ``effective'' source
redshift by requiring that the angular diameter distance to the source
redshift is twice as large as that to the lens redshift, i.e.,
$d_A(0;z_S)=2d_A(0;z_L)$. With this requirement, the source redshifts are
$z_s=0.65$, $1.19$, and $2.25$ for $z_L=0.3$, $0.5$, and $0.8$, respectively.

Figure~\ref{fig:cl_gglens_diff} shows the fractional differences between
non-Gaussian predictions and the Gaussian prediction ($\fnl=0$), which
are simply equal
to $\Delta b(k,z_L)/b_1(z_L)$ where $k=l/d_A(0;z_L)$.
As expected from the form of the scale-dependent bias, the difference
grows toward small multipoles as roughly $1/l^2$.
While lower redshift populations do not show more than 10\% difference
at $l\ge 10$ for $\fnl=\pm 50$,
a higher-$z$ population of lens galaxies or clusters of galaxies
at $z_L=0.8$ show the differences at the level of $\sim 10$\% at $l\sim 20$
and $\sim 30\%$ at $l\sim 10$. Are these effects detectable?

\subsection{Covariance matrix of the galaxy-convergence cross power
  spectrum}
\label{sec:Clgk_cov}
\begin{figure}[t]
\begin{center}
\includegraphics[width=8cm]{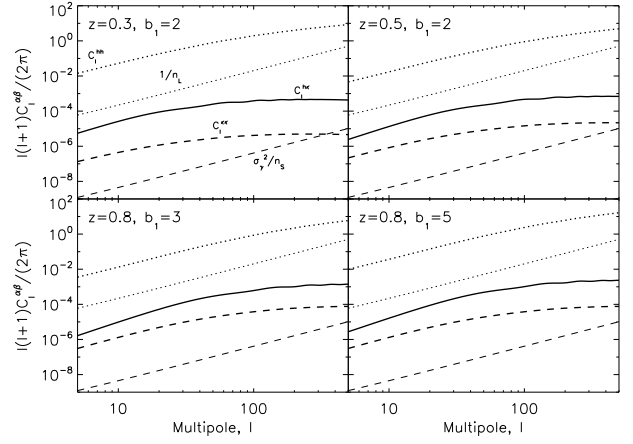}
\end{center}
\caption{
Angular power spectra of the galaxy-galaxy correlation, $C_l^{h}$
(thick dotted lines), the galaxy-convergence cross-correlation, $C_l^{h\kappa}$ (thick solid lines), and the
convergence-convergence correlation, $C_l^{\kappa}$ (thick dashed lines)
 for the Gaussian initial condition ($\fnl=0$).
The four panels show the same populations of galaxies and clusters of
 galaxies as in Fig.~\ref{fig:cl_gglens_signal}.
We also show the galaxy shot noise, $1/n_L$ (thin dotted lines) as well
 as the source shape noise,
$\sigma^2_\gamma/n_S$  (thin dashed lines), for $N_L=10^6$,
 $\sigma_\gamma=0.3$,  and $n_S=3.5\times10^{8}~\mathrm{sr}^{-1}$.
We find $1/n_L\ll C_l^h$ and $\sigma_\gamma^2/n_S\ll C_l^\kappa$ for
 $l\lesssim 100$.
\label{fig:Cls_show}
}
\end{figure}
\begin{figure}[t]
\begin{center}
\includegraphics[width=8cm]{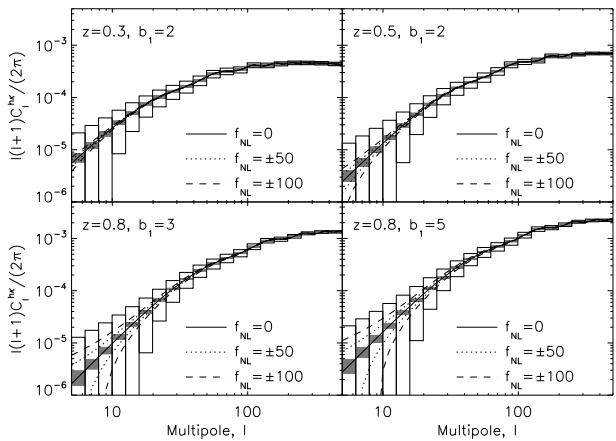}
\end{center}
\caption{
Same as Fig.~\ref{fig:cl_gglens_signal},
with the expected 1-$\sigma$
 uncertainties for full-sky lens surveys
and a single lens redshift. Adjacent bins are uncorrelated.
The open (filled) boxes show the binned uncertainties
 with (without) the cosmic variance term due to the cosmic shear field
 included. We used Eq.~(\ref{eq:gglens_cl_cov}) for the open boxes, and
 Eq.~(\ref{eq:gglens_cl_cov}) with $C_l^{h\kappa}=0=C_l^\kappa$
 for the filled boxes. We use logarithmic bins of $\Delta l=0.23 l$.
Note that the uncertainties are calculated for a single lens redshift
 slice, and thus they will go down as we add more lens redshift slices.
\label{fig:cl_gglens_err}
}
\end{figure}

The covariance matrix of the galaxy-convergence cross-correlation power
spectrum is given by
\begin{eqnarray}
\nonumber
&&
\langle C_{l}^{h\kappa} C_{l'}^{h\kappa} \rangle
-\langle C_{l}^{h\kappa} \rangle\langle C_{l'}^{h\kappa} \rangle
\\
&=&
\frac{\delta_{ll'}}
{(2l+1)f_\mathrm{sky}}
\left[
\left(C_l^{h\kappa}\right)^2
+
\left(C_l^h+\frac{1}{n_L}\right)
\left(C_l^\kappa+\frac{\sigma_{\gamma}^2}{n_S}\right)
\right],
\label{eq:gglens_cl_cov}
\end{eqnarray}
where $\delta_{ll'}$ is Kronecker's delta symbol showing that
the angular power spectra at different multipoles are uncorrelated.
Again, $C_l^{h}$ and $C_l^{\kappa}$ are the angular
power spectra of the lens halos (galaxies or cluster of galaxies) and $\kappa$, respectively, and $n_L$ and $n_S$ are the number densities of the lens halos
and the lensed (source) galaxies, respectively.

We calculate $C_l^{\kappa}$ by using Limber's approximation as
\begin{eqnarray}
C_l^\kappa
=\int_0^{z_S} dz
\frac{\rho_0^2}{\Sigma_c^2(z;z_S)}
\frac{P_m\left[k=\frac{l+1/2}{d_A(0;z)};z\right]}{H(z)d_A^2(0;z)}.
\label{eq:Clkk_limber}
\end{eqnarray}
However, we cannot use Limber's approximation for $C_l^h$ unless one
considers lens redshift slices that are broad. As we are assuming a thin
lens redshift slice throughout this paper, we must not use Limber's
approximation, but
evaluate the exact
integral relation:
\begin{equation}
C_l^h
=\frac{2}{\pi}\int dk k^2P_g(k,z_L) j_l^2\left[kd_A(z_L)\right],
\label{eq:Clgg_exact}
\end{equation}
where $j_l$ is the spherical Bessel function, and
$P_g(k,z)$ is the linear galaxy power spectrum: $P_g(k)=b^2_1P_m(k)$.

Fig.~\ref{fig:Cls_show} shows the galaxy-galaxy, galaxy-convergence, and
convergence-convergence angular power spectra for Gaussian ($\fnl=0$)
initial conditions. We also show the shot noise
of the galaxy angular power spectrum, $1/n_L$, and the shape noise of
the convergence power spectrum, $\sigma_\gamma^2/n_S$, with the
following representative values: $N_L=4\pi n_L = 10^6$, $n_S=3.5\times
10^8~\mathrm{sr}^{-1}$,  and
$\sigma_\gamma=0.3$.
We find $1/n_L\ll C_l^h$ and $\sigma_\gamma^2/n_S\ll C_l^\kappa$ for
the multipoles that we are interested in, i.e., $l\lesssim 100$, and
thus we conclude that the uncertainties are totally dominated by the
cosmic variance terms. In other words, the size of the uncertainties are
insensitive to the exact choices of $N_L$, $\sigma_\gamma$, or $n_S$.

We also find that the values of cross correlation coefficients,
$r_l\equiv C_l^{h\kappa}/\sqrt{C_l^h C_l^\kappa}$, are small (of order
10--20\%): the maximum values are $0.19$, $0.15$, and $0.13$ for
$z_L=0.3$, $0.5$, and $0.8$, respectively.
This implies that one may ignore the contribution of $C_l^{h\kappa}$ to
the covariance matrix, approximating the variance of $C_l^{h\kappa}$ of
a single lens redshift slice for a multipole bin of size
 $\Delta l$ as:
\begin{equation}
\mathrm{Var}(C_l^{h\kappa})
=
\frac{C_l^hC_l^{\kappa}}{(2l+1)\Delta lf_\mathrm{sky}}.
\label{eq:var_cv_dominate}
\end{equation}
Therefore, we should be able to measure the galaxy-convergence
cross-power spectrum with $C_l^{h\kappa}/\sqrt{{\rm Var}(C_l^{h\kappa})}\gtrsim
1$ when the multipoles satisfy
\begin{equation}
l \gtrsim l_{\rm min}\equiv
\frac{1}{r_l\sqrt{2(\Delta l/l)f_\mathrm{sky}}}.
\label{eq:find_lmin}
\end{equation}
For the galaxy-convergence power spectra in Fig.~\ref{fig:cl_gglens_signal}
with the full sky coverage ($f_\mathrm{sky}=1$) and $\Delta l/l=0.23$,
we find $l_{min}=9.0$, $12.1$, and $15.7$
for $z_L=0.3$ ($z_S=0.65$), $0.5$ ($1.19$), and $0.8$ ($2.25$), respectively.

Similarly, we can estimate the maximum radius below which
we can measure the mean tangential shear, $\Delta\Sigma(R)$, as
\begin{equation}
R_\mathrm{max}\simeq\frac{\pi d_A(0;z_L)}{l_\mathrm{min}}.
\end{equation}
For example, with $\Delta R/R=\Delta l/l=0.1$, we get
$R_\mathrm{max}\simeq 215$, $260$, and $300~h^{-1}~\mathrm{Mpc}$ for
$z_L=0.3$, $0.5$ and $0.8$, respectively. These values do give the radii
at which the signal-to-noise ratios are roughly unity in
Fig.~\ref{fig:Dsigma_ebar}.

Fig.~\ref{fig:cl_gglens_err} shows the expected 1-$\sigma$ uncertainties
of $C_l^{h\kappa}$ for several populations of lens galaxies.
We find that the cosmic variance completely dominates the uncertainties
on large scales (low $l$) where the non-Gaussian effects are the
largest. Again, while we find that it would be difficult to measure $\fnl$ from
a {\em single lens redshift slice}, combining many redshift slices
should help us measure $\fnl$, especially when we can use many slices at
moderately high redshifts.

\section{Halo-mass correlation from galaxy-CMB lensing}
\label{sec:cmblens}
\subsection{Formula}
Instead of using the background galaxies for measuring the cosmic shear
field due to the intervening mass, one can use the CMB as the background
light and measure the shear field of the {\it CMB lensing} due to the
intervening mass between us and the the photon decoupling epoch at
$z_*\simeq 1089$. See \cite{lewis/challinor:2006} for a review on the
CMB lensing.

The lensing effect makes CMB anisotropies (both temperature
and polarization) non-Gaussian by producing a non-vanishing connected
four-point function, although it does not produce any non-vanishing
three-point function. One can use this property to reconstruct the
lensing potential field, hence the projected mass-density field between
us and $z_*$, from the four-point function of CMB
\cite{hu/okamoto:2002,okamoto/hu:2003,hirata/seljak:2003}.

By cross-correlating the halo over-density field, $\delta_h$, at some
redshift $z_L$ (measured from spectroscopic observations) and the $\kappa$
field reconstructed from the CMB lensing, one can measure the halo-convergence
angular power spectrum, $C_l^{h\kappa}$.

The angular power spectrum of the galaxy-CMB lensing cross correlation
is merely a special case of the galaxy-convergence cross correlation
that we have studied
in the previous section: all we need to do is to set the source
redshift, $z_S$, to be the redshift of the photon decoupling epoch,
$z_*\simeq 1089$, i.e., $z_S=z_*$.
Note that for a flat universe $d_A(z_L;z_*)=d_A(0;z_*)-d_A(0;z_L)$ where
$d_A(0;z_*)=9.83~h^{-1}~{\rm Gpc}$.

\begin{figure}[t]
\begin{center}
\includegraphics[width=9cm]{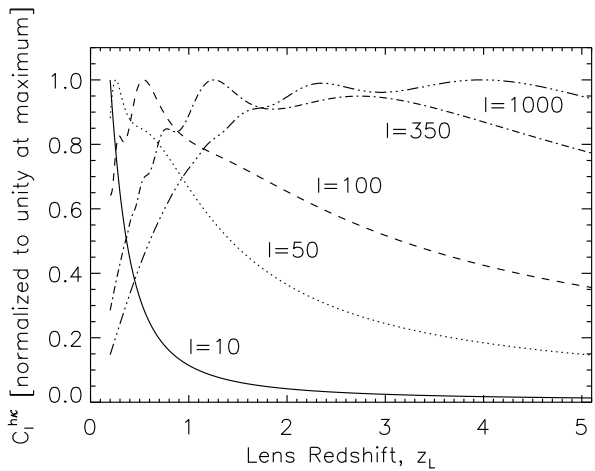}
\end{center}
\caption{
\label{fig:cl_cmb_redshift}
Angular power spectrum of the galaxy-CMB lensing, $C_l^{h\kappa}$, at
 various multipoles as a function of the lens redshift, $z_L$. We have
 divided $C_l^{h\kappa}$ by its maximum value. The solid, dotted,
 dashed, dot-dashed, and triple-dot-dashed lines show $l=10$, 50, 100,
 350, and 1000, respectively.
}
\end{figure}

Figure~\ref{fig:cl_cmb_redshift} shows that the CMB lensing at low
(high) multipoles are better correlated with low-$z$ (high-$z$)
galaxies. This is due to the shape of the matter power spectrum, as we have
explained in the previous section. Note that $C_l^{h\kappa}$ of the CMB
lensing for a given
multipole decreases more slowly with $z_L$  than that of the galaxy
lensing due to  the geometrical factor $d_A(z_L;z_S)/d_A(0;z_S)$.

Note that CMB and galaxies at $z\lesssim 1$ are correlated also via the
Integrated Sachs-Wolfe (ISW) effect
\citep{boughn/crittenden:2004}.
We shall not include this effect in our cross-correlation calculation
for the following reason.
We calculate the cross-correlation signal between galaxies and the
convergence field reconstructed from CMB. This reconstruction relies on
the fact that lensed CMB fluctuations have non-vanishing connected
four-point function. On the other hand, the linear ISW effect does not
have such a particular form of four-point function induced by lensing,
and thus should not contribute to the reconstructed convergence field.
See \cite{afshordi/tolley:2008} for the effects of $\fnl$ on the
galaxy-ISW cross correlation.

\subsection{Results}
We can now calculate $C_l^{h\kappa}$ for various values of $\fnl$. We
use
\begin{eqnarray}
\nonumber
 C_l^{h\kappa}(z_L)
&=& \frac{4\pi G\rho_0}{c^2}(1+z_L)\frac{d_A(z_L;z_*)}{d_A(0;z_L)d_A(0;z_*)}\\
\nonumber
& &\times
\left[b_1(z_L)+\Delta b\left(k=\frac{l}{d_A(0;z_L)},z_L\right)\right]\\
& &\times
P_{m}\left[k=\frac{l}{d_A(0;z_L)},z_L\right],
\end{eqnarray}
where the scale-dependent bias, $\Delta b(k,z)$, is given by
Eq.~(\ref{eq:bk}).

\begin{figure}[t]
\begin{center}
\includegraphics[width=8cm]{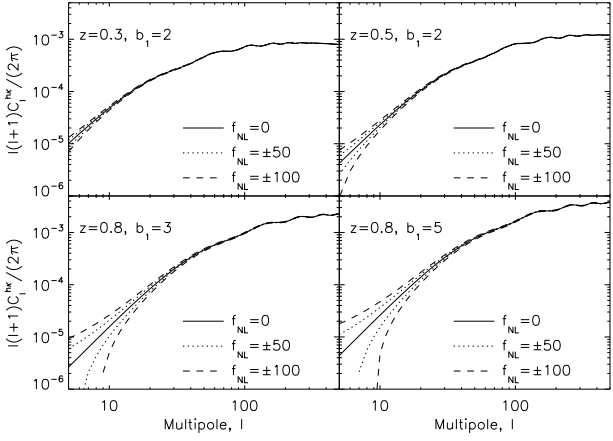}
\end{center}
\caption{
\label{fig:cl_cmb_lowz}
Imprints of the local-type primordial non-Gaussianity in the
 galaxy-CMB lensing power spectrum, $l(l+1)C_l^{h\kappa}/(2\pi)$,
for the same populations of
 lens galaxies as in Fig.~\ref{fig:bao}.
 The solid, dashed, and dotted lines show $\fnl=0$, $\pm 50$, and $\pm
 100$, respectively.
}
\end{figure}

Figure~\ref{fig:cl_cmb_lowz} shows $C_l^{h\kappa}(z_L)$ for $\fnl=\pm
50$ and $\pm 100$ for populations of low-$z$ galaxies that we have
considered in the previous sections: $b_1=2$ at $z_L=0.3$ (similar to
SDSS LRGs, top-left),  $b_1=2$ at $z_L=0.5$ (higher-$z$ LRGs, top-right),
$b_1=2$ at $z_L=0.8$ (galaxies that can be observed by LSST, bottom-left),
and $b_1=5$ at $z_L=0.8$ (clusters of galaxies that can be observed by LSST,
bottom-right). The fractional differences between non-Gaussian
predictions and  the
Gaussian prediction ($\fnl=0$) are exactly the same as those shown in
Fig.~\ref{fig:cl_gglens_diff}: in the limit where Limber's approximation
is valid, the
galaxy-convergence power spectrum and the galaxy-CMB lensing power spectrum
for the same lens galaxies differ only by a constant geometrical factor of
$d_A(z_L;z_*)d_A(0;z_S)/d_A(z_L;z_S)d_A(0;z_*)$.
Incidentally, for our choice of the source redshifts in the previous section,
$2d_A(z_L;z_*)/d_A(0;z_*)=1.83$, $1.73$, and $1.60$
for $z_L=0.3$, $0.5$, and $0.8$, respectively.

Therefore, the galaxy-CMB lensing cross correlation would
provide a nice cross-check for systematics of the galaxy-convergence
cross correlation, and \textit{vice versa}: after all, we are measuring
the same quantity, $P_{hm}(k)$,
by two different background sources, high-z galaxies and CMB.

In using high-$z$ galaxies as sources, the galaxy-galaxy lensing
measurement may be susceptible to systematic errors widely discussed in
the lensing literature, namely shear calibration, coherent point spread
function (PSF) anisotropy, redshift biases, magnification bias and
intrinsic alignments of galaxies. Here we are particularly concerned
with errors that affect galaxy-shear cross-correlations by mimicking the
angular dependence of the signal due to non-zero $\fnl$. Fortunately
most systematic errors that affect shear-shear correlations do not
contribute to galaxy-shear cross correlations: for instance, PSF anisotropy
affects background galaxy shapes but not foreground galaxy locations
\citep{mandelbaum/etal:2005}. With standard lensing
data analysis methods, it can be ensured that both the shear calibration
and PSF do not contribute a scale dependence to the first order. Biases in
the redshift distributions of lens and source galaxies can similarly
lead to a mis-estimation of the amplitude of the signal, but not its
scale dependence. Thus, to the lowest order, the measurement of $\fnl$ via the
scale dependence of the galaxy-galaxy lensing signal is robust to the
leading systematic errors in weak lensing. But a detailed study of
various sources of error is needed given the small signal we are
seeking.

\begin{figure}[t]
\begin{center}
\includegraphics[width=8cm]{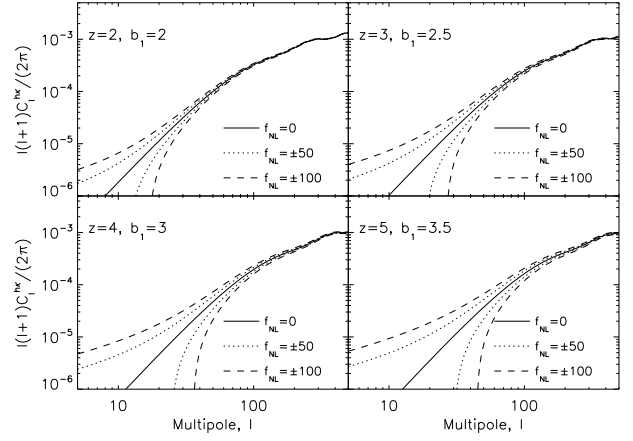}
\end{center}
\caption{
\label{fig:cl_cmb_highz}
Same as Fig.~\ref{fig:cl_cmb_lowz}, but for
high-$z$ lens galaxies with
 $b_1=2$ at $z_L=2$ (top-left),
$b_1=2.5$ at $z_L=3$ (top-right),
$b_1=3$ at $z_L=4$ (bottom-left),
and $b_1=3.5$ at $z_L=5$ (bottom-right).
}
\end{figure}

\begin{figure}
\begin{center}
\includegraphics[width=8cm]{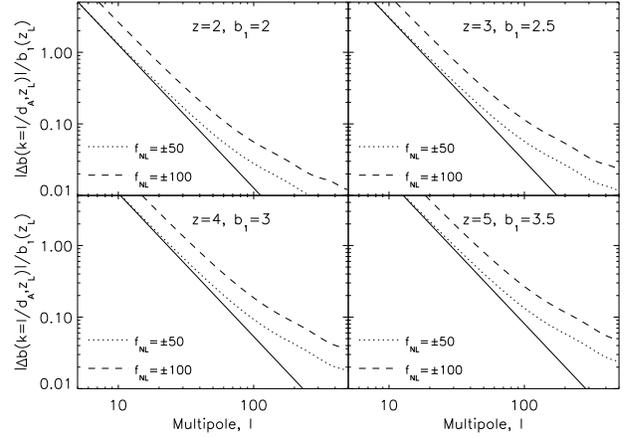}
\end{center}
\caption{
\label{fig:cl_cmb_diff_highz}
Same as Fig.~\ref{fig:cl_gglens_diff}, but for
high-$z$ lens galaxies with
 $b_1=2$ at $z_L=2$ (top-left),
$b_1=2.5$ at $z_L=3$ (top-right),
$b_1=3$ at $z_L=4$ (bottom-left),
and $b_1=3.5$ at $z_L=5$ (bottom-right).
}
\end{figure}

Another benefit of using the CMB lensing as a proxy for the intervening matter
distribution is that we can probe the galaxy-matter cross correlation at high
redshift to which we cannot reach with the galaxy-galaxy lensing method.
It is especially useful for probing primordial non-Gaussianity, as the
scale dependent bias
signal is higher for higher lens redshift: $\Delta b(k,z_L)\propto 1/D(z_L)$
(see Eq.~(\ref{eq:bk})).
Therefore, we find that even higher-$z$ populations of galaxies give us a much
better chance of detecting the effects of $\fnl$. Figure~\ref{fig:cl_cmb_highz} shows $C_l^{h\kappa}(z_L)$ for $\fnl=\pm
50$ and $\pm 100$ for populations of high-$z$ galaxies:
$b_1=2$ at $z_L=2$ (top-left),
$b_1=2.5$ at $z_L=3$ (top-right),
$b_1=3$ at $z_L=4$ (bottom-left),
and $b_1=3.5$ at $z_L=5$ (bottom-right).
The first one, a spectroscopic galaxy survey at $z_L=2$ with $b_L=2$, is
within reach by, e.g., the Hobby-Eberly Telescope Dark Energy Experiment
(HETDEX) \cite{hill/etal:2004,hill/etal:prep}.
There we find, for $\fnl=\pm 50$, $\sim 10$\% effect at $l\sim 40$, and
a factor of two effect at $l\sim 10$
(see Fig.~\ref{fig:cl_cmb_diff_highz}).
The effects grow bigger at higher
$z$: higher-$z$ surveys at
$z>3$ can be done with, e.g., the concept of the Cosmic Inflation Probe (CIP)
\footnote{http://www.cfa.harvard.edu/cip/}.
At $z_L=4$ and 5 (with $b_1=3$ and 4, respectively) we find $\sim 10$\%
effect at $l\sim 100$, a factor of two effect at $l\sim 30$, and even
bigger effects at $l\lesssim 30$
(see Fig.~\ref{fig:cl_cmb_diff_highz}).


\subsection{Covariance matrix of the galaxy-CMB lensing}
\begin{figure}
\begin{center}
\includegraphics[width=8cm]{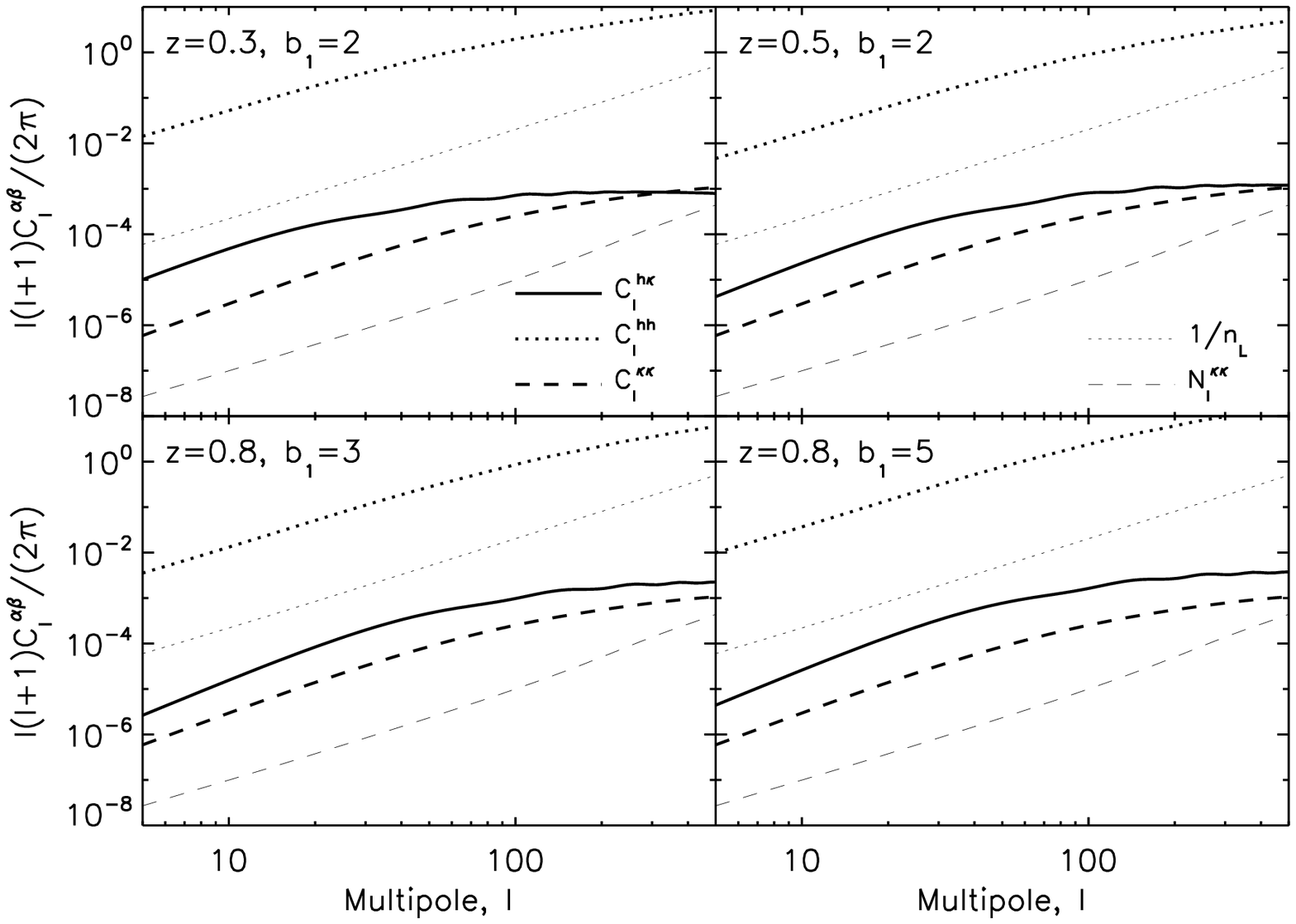}
\end{center}
\caption{
\label{fig:cls_cmb_show_lowz}
Angular power spectra of the galaxy-galaxy correlation, $C_l^{h}$
(thick dotted lines), the galaxy-convergence cross-correlation, $C_l^{h\kappa}$ (thick solid lines), and the
convergence-convergence correlation, $C_l^{\kappa}$ (thick dashed lines)
 for the Gaussian initial condition ($\fnl=0$).
The four panels show the same populations of galaxies and clusters of
 galaxies as in Fig.~\ref{fig:cl_cmb_lowz}.
We also show the galaxy shot noise, $1/n_L$ (thin dotted lines) as well
 as the lens reconstruction noise, $N_l^\kappa$ (think dashed lines),
for $N_L=10^6$ and $N_l^\kappa\simeq 6\times10^{-8}$ $\mathrm{sr}^{-1}$
 (for multipoles much smaller than that corresponds to the beam size of $4'$).
We find $1/n_L\ll C_l^h$ and $N_l^\kappa\ll C_l^\kappa$ for
 $l\lesssim 100$.
}
\end{figure}
\begin{figure}[t]
\begin{center}
\includegraphics[width=8cm]{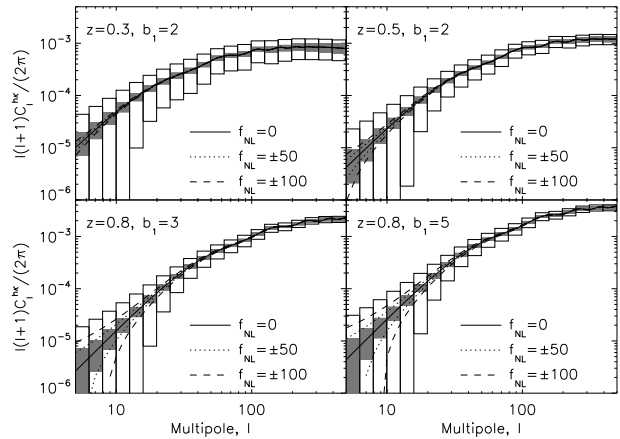}
\end{center}
\caption{
\label{fig:cl_cmb_lowz_err}
Same as Fig.~\ref{fig:cl_cmb_lowz},
but with 1-sigma uncertainty due to the
shape noise of source galaxies (filled box, Eq.~(\ref{eq:binned_err_shape}))
and full error budget (empty box, diagonal of Eq.~(\ref{eq:binned_cov}))
including the cosmic variance.
We use the multipole bins of size $\Delta l=0.23 l$.
For uncertainty of CMB lensing reconstruction,
We assume the nearly-perfect reference experiment of \citet{hu/okamoto:2002}:
white detector noise $\Delta_T=\Delta_P/\sqrt{2}=1~\mu K~\mathrm{arcmin}$,
and FWHM of the beam $\sigma=4'$.
}
\end{figure}
\begin{figure}
\begin{center}
\includegraphics[width=8cm]{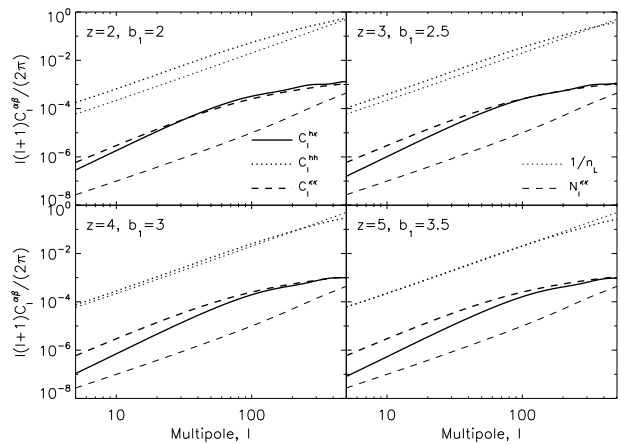}
\end{center}
\caption{
\label{fig:cls_cmb_show_highz}
Same as Fig.~\ref{fig:cls_cmb_show_lowz}, but for the high redshift
lens galaxies shown in Fig.~\ref{fig:cl_cmb_highz}.
For these
 populations (and with $N_L=10^6$), the shot noise is about the same as
 the galaxy power spectrum, i.e., $C_l^h\simeq 1/n_L$.
}
\end{figure}
\begin{figure}
\begin{center}
\includegraphics[width=8cm]{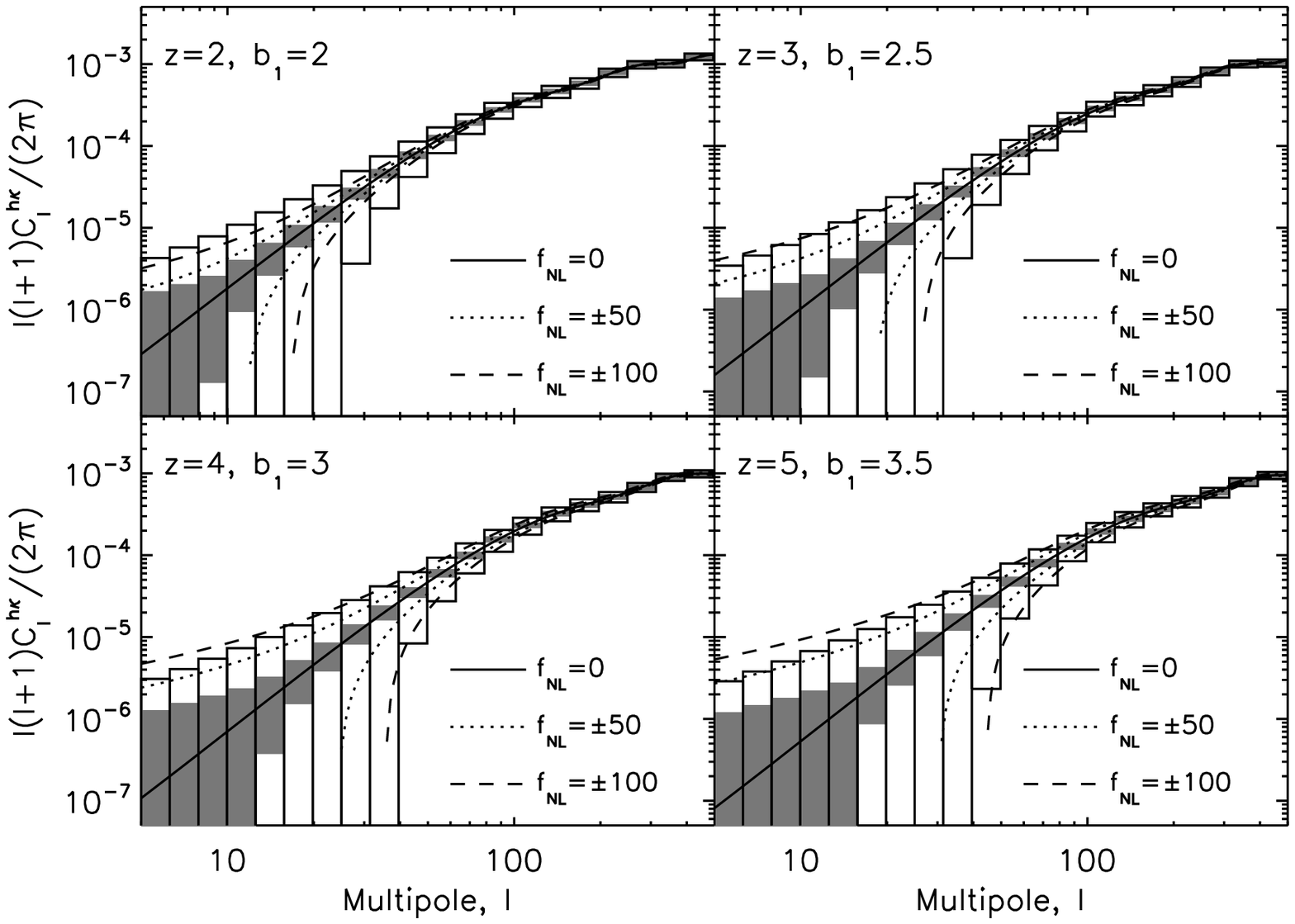}
\end{center}
\caption{
\label{fig:cl_cmb_highz_err}
Same as Fig.~\ref{fig:cl_cmb_lowz_err}, but for the high redshift
lens galaxies shown in Fig.~\ref{fig:cl_cmb_highz}.
}
\end{figure}
The covariance matrix of the galaxy-CMB lensing is given by
\cite{acquaviva/etal:2008}
\begin{eqnarray}
\nonumber
& & \langle C_l^{h\kappa}C_{l'}^{h\kappa}\rangle
-
\langle C_l^{h\kappa}\rangle\langle C_{l'}^{h\kappa}\rangle\\
&=&
\frac{(C_l^{h\kappa})^2+\left(C_l^{h}+1/{n_L}\right)\left(C_l^{\kappa}+N_l^{\kappa}\right)}{(2l+1)f_{\rm
sky}}\delta_{ll'},
\end{eqnarray}
where $N_l^{\kappa}$ is the reconstruction noise from CMB given by
\cite{hu/okamoto:2002}.
The covariance matrix equation here is the same as
Eq.~(\ref{eq:gglens_cl_cov}), except that now the shape noise of source
galaxies is replaced by the reconstruction noise of CMB lensing.
In what follows, we shall assume a ``nearly perfect''
CMB experiment considered in \citet{hu/okamoto:2002}, whose Gaussian random
detector noise is modeled as \citep{knox:1995}
\begin{eqnarray}
\nonumber
C_l^{T}\biggl|_\mathrm{noise}
&=&
\left(\frac{T_\mathrm{CMB}}{\Delta_T}\right)^{-2} e^{l(l+1)\sigma^2/8 \ln 2}
,
\\
C_l^{E}\biggl|_\mathrm{noise}
=
C_l^{B}\biggl|_\mathrm{noise}
&=&
\left(\frac{T_\mathrm{CMB}}{\Delta_T}\right)^{-2} e^{l(l+1)\sigma^2/8 \ln 2}
,
\end{eqnarray}
where the white noise level of detectors is
$\Delta_T=\Delta_P/\sqrt{2}=1~\mu K~\mathrm{arcmin}$,
and the Full-Width-at-Half-Maximum (FWHM) of the beam is $\sigma=4'$.
With these detector parameters and the cosmological parameters of the
``WMAP+BAO+SN ML'' parameters in Table 1 of \cite{komatsu/etal:2009},
we find $N_l^\kappa\simeq 6\times10^{-8}$ $\mathrm{sr}^{-1}$
on large scales, $l<100$.

Fig.~\ref{fig:cls_cmb_show_lowz} shows the galaxy-galaxy, galaxy-convergence,
convergence-convergence angular power spectra for the Gaussian initial
condition ($\fnl=0$).
This figure is qualitatively similar to Fig.~\ref{fig:Cls_show}:
the galaxy-galaxy correlation is exactly the same, and the galaxy-convergence
power spectrum is simply a scaled version of the corresponding curve
in Fig.~\ref{fig:Cls_show}.
The major difference comes from $C_l^\kappa$: as the CMB photons travel
a longer path than photons from source galaxies,
the convergence-convergence power spectrum is higher for the CMB
lensing convergence.

On large scales ($l\lesssim 100$), the covariance matrix is dominated by
the cosmic variance terms:  $1/n_L\ll C_l^h$ and $N_l^\kappa\ll C_l^\kappa$.
The cross correlation coefficients are small, of order 10\%:
the maximum values are $0.12$, $0.11$, and $0.10$ for $z_L=0.3$, $0.5$,
and $0.8$, respectively. Therefore, we can again use
Eq.~(\ref{eq:var_cv_dominate})  for estimating the variance, and find
$l_\mathrm{min}$  (Eq.~(\ref{eq:find_lmin}))
above which we can measure the galaxy-convergence cross correlation with
the signal-to-noise ratio greater than unity.
For logarithmic bins of $\Delta l/l=0.23$,
we find $l_\mathrm{min}=12.2$, $13.5$, and $15.8$ for $z_L=0.3$, $0.5$,
and $0.8$, respectively.
Comparing to the results in Sec~\ref{sec:Clgk_cov}, $l_\mathrm{min}$ is
slightly bigger, as $C_l^\kappa$ (which contributes to the uncertainty)
increases more rapidly than $C_l^{h\kappa}$ (the signal we are after) would
as the source redshift increases from $z_S$ to $z_*$.

Fig.~\ref{fig:cl_cmb_lowz_err} shows the expected 1-$\sigma$
uncertainties of
the angular power spectrum of the galaxy-CMB lensing cross correlation,
on top of the predicted Gaussian/non-Gaussian signals with five
different values of
non-Gaussianity parameters: $\fnl=0$, $\pm 50$, $\pm 100$.
We also show the 1-$\sigma$ uncertainties without the cosmic variance
due to the cosmic shear. Once again, it would be difficult to measure
the effects of $\fnl$ from a single lens redshift, but combining many
slices would help measure $\fnl$ from the galaxy-CMB lensing cross
correlation.

What about using even higher-$z$ lens galaxies? As shown in
Fig.~\ref{fig:cls_cmb_show_highz}, for higher-$z$ populations (with
$z_L=2-5$)
the galaxy-galaxy power spectra are about the same as
the shot noise levels. This is true only for the assumed number of
lenses, $N_L=10^6$ (over the full sky), which is somewhat
arbitrary. Increasing $N_L$ will
help reduce the noise, but only up to a factor of $\sqrt{2}$.
For populations with $C_l^h\simeq 1/n_L$, we can approximate the variance as
\begin{equation}
\mathrm{Var}(C_l^{h\kappa})
=
\frac{(C_l^h+1/n_L)C_l^\kappa}{(2l+1)\Delta lf_\mathrm{sky}}
\simeq
\frac{2C_l^hC_l^\kappa}{(2l+1)\Delta lf_\mathrm{sky}}.
\end{equation}
Thus, we find $C_l^{h\kappa}/\sqrt{\mathrm{Var}(C_l^{h\kappa})}\gtrsim
1$ when
\begin{equation}
l \gtrsim \frac{1}{r_l\sqrt{(\Delta l/l)f_\mathrm{sky}}}.
\label{eq:find_lmin_highz}
\end{equation}
The maximum cross-correlation coefficients are
$0.091$, $0.084$, $0.078$, and $0.073$ for
$z_L=2$, $3$, $4$, and $5$, respectively.
The estimated $l_\mathrm{max}$ is then $29$ ($z_L=2$), $34$ ($z_L=3$), $38$ ($z_L=4$) and $42$ ($z_L=5$).

In Fig.~\ref{fig:cl_cmb_highz_err} we compare  the expected 1-$\sigma$
uncertainties with the predicted signals from high-$z$ lens galaxies with
$\fnl=0$, $\pm 50$, and $\pm 100$. Comparing this result with that in
Fig.~\ref{fig:cl_cmb_lowz_err}, we conclude that higher-$z$ lens
populations do provide a better chance of finding the effects of $\fnl$
than lower-$z$ lenses, although we would still need to combine  many
lens redshift slices. In particular, using higher-$z$ lenses, we can
find non-Gaussian effects at higher and higher multipoles which are
easier to measure; thus, high-$z$ galaxies correlated with CMB lensing
offers a yet another nice probe of the local-type primordial non-Gaussianity.

\section{Discussion and Conclusions}

In this paper we have studied the galaxy-galaxy lensing and galaxy-CMB
lensing cross-correlation functions. We have focused on large scales,
typically larger than 100~Mpc at the lens redshift. While current
measurements have high signal-to-noise ratios on much smaller scales, we
believe that future surveys will enable detection of interesting physical
effects in the large-scale, linear regime.

We derive the full covariance matrix for galaxy-galaxy lensing,
including the cosmic variance due to the clustering of lenses and to
cosmic shear (Eq.~\ref{eq:cov}). We use the linear bias model to provide the
halo-mass
and halo-halo correlations needed for this calculation. We present
results for the covariance of the mean tangential shear measurement as a
function of angular separations, as well as for the harmonic space
halo-convergence cross-power spectrum.
Our calculations show
that the errors in
       $\Delta\Sigma(R)$ are dominated by the cosmic variance term for
       $R\gtrsim 50~h^{-1}~{\rm Mpc}$
       (see Fig.~\ref{fig:bao_ebar}).
       Similarly, the errors in the
       halo-convergence cross power spectra, $C_l^{h\kappa}$, are
       dominated by the cosmic variance term at $l\lesssim 100$
       (see Fig.~\ref{fig:cl_cmb_lowz_err}).

For Gaussian initial conditions, we show that
the baryonic effects in the matter power spectrum
       (often called Baryon Acoustic Oscillations) produce a
       ``shoulder'' in the galaxy-galaxy lensing correlation (i.e., the mean
       tangential shears),
       $\Delta\Sigma(R)$, at $R\sim 110~h^{-1}~{\rm Mpc}$
       (see Fig.~\ref{fig:bao}).
       This effect
       should be easy to measure from the next-generation lensing
       surveys by combining $\Delta\Sigma(R)$ from multiple
       lens redshift slices.

We consider the prospects of detecting primordial non-Gaussianity of the
local-form, characterized by the $\fnl$ parameter.
We have found that
the scale-dependent bias from the local-form non-Gaussianity with
       $\fnl=\pm 50$ modifies $\Delta\Sigma(R)$ at the level of 10--20\%
       at $R\sim 300~h^{-1}~{\rm Mpc}$ (depending on $b_1$ and $z_L$; see
Fig.~\ref{fig:fnl_gglens_diff})
(see Fig.~\ref{fig:fnl_gglens_signal}).
The modification grows rapidly toward
larger scales, in proportion to $R^2$.
High-$z$ galaxies at, e.g., $z\gtrsim 2$, cross-correlated with
       CMB can be used to find the effects of $\fnl$ in the
       galaxy-convergence power spectrum, $C_l^{h\kappa}$. While the effects
       are probably too small to see from a single lens redshift
       (see Fig.~\ref{fig:cl_cmb_highz_err}),
       many
       slices can be combined to beat down the cosmic variance
       errors. Exactly how many slices are necessary, or what is the
       optimal strategy to measure $\fnl$ from the galaxy-CMB lensing
       signal requires a more detailed study that incorporates
       the survey strategy for specific galaxy and lens surveys.

We emphasize that, while the two-point statistics of shear fields are
not sensitive
to primordial non-Gaussianity, the two-point statistics correlating shear
fields with density peaks (i.e., galaxies and clusters of galaxies) are
sensitive 
due to the strong scale-dependence of halo bias on large scales.

Finally, we note that one can
also measure the effects of $\fnl$ on the halo power spectrum,
$C_l^h$. For example, $C_l^h$ that would be measured from LSST can be
used to probe $\fnl\sim 1$ \citep{carbone/verde/matarrese:2008}; thus,
we would expect $C_l^h$ to be more powerful than the lens cross-correlation
statistics we studied here.
However, a combination of the two measurements would provide
useful cross-checks,
as galaxy clustering and galaxy-lensing correlations are affected by
very different systematics.

\acknowledgments
E.K. would like to thank Erin Sheldon and Rachel Mandelbaum for very
useful discussions.
B.J. thanks the UT Austin astronomy department and IUCAA for their
hospitality while part of this work was carried out, and Gary Bernstein,
Sarah Bridle, and Ishaana Monet for stimulating discussions.
This material is based in part upon work
supported by the Texas Advanced Research Program under
Grant No. 003658-0005-2006, by NASA grants NNX08AM29G
and NNX08AL43G, and by NSF grants AST-0807649 and PHY-0758153.
E.K. acknowledges support from an Alfred P. Sloan Fellowship.
D.J. acknowledges support from a  Wendell Gordon Endowed Graduate
Fellowship of the University of Texas at Austin.
B.J. is partially supported by NSF grant AST-0908027.

\appendix
\section{Derivation of the  mean tangential
 shear}
\label{app:meanshear}

One may write down the observed tangential shears
at a given distance from a lens halo, $\bm{\theta}$, averaged over $N_L$ lens halos as
\begin{equation}
 {\gamma}_t^h(\bm{\theta})=
\frac1{N_L}\int d^2\hat{\mathbf n}
\left[\sum_i^{N_L}
\delta_D(\hat{\mathbf
n}-\hat{\mathbf n}_i)\right]\gamma_{t}(\hat{\mathbf
n}+\bm{\theta}),
\end{equation}
where $\delta_D$ is the delta function, and $i$ denotes the location of
lens halos.
Note that we have not azimuthally
averaged the tangential shears yet.
The ensemble average of $\gamma_t^h$ yields the number-weighted average
of the tangential shear:
\begin{equation}
 \langle\gamma_t^h\rangle(\bm{\theta})
= \frac1{N_L}\int d^2\hat{\mathbf n}\langle n_L(\hat{\mathbf n})
\gamma_{t}(\hat{\mathbf n}+\bm{\theta})\rangle,
\end{equation}
where $n_L(\hat{\mathbf n})$ is the surface number density of lens halos at a
given location on the sky, $\hat{\mathbf n}$. Expanding it into the
perturbation, $n_L(\hat{\mathbf n})=\bar{n}_L[1+\delta_h(\hat{\mathbf
n})]$, we obtain
\begin{equation}
 \langle\gamma_t^h\rangle(\bm{\theta})
= \frac1{f_{\rm sky}}\int \frac{d^2\hat{\mathbf n}}{4\pi}\langle \delta_h(\hat{\mathbf n}) \gamma_{t}(\hat{\mathbf n}+\bm{\theta})\rangle,
\end{equation}
where $f_{\rm sky}\equiv N_L/(4\pi\bar{n}_L)$ is a fraction of sky covered by
the observation.
From statistical isotropy of the universe, $\langle
\delta_h(\hat{\mathbf n}) \gamma_{t}(\hat{\mathbf
n}+\bm{\theta})\rangle$ does not depend on $\hat{\mathbf n}$, and thus
the integral over $\hat{\mathbf n}$ simply gives $4\pi f_{\rm sky}$.
Expanding $\delta_h$ and $\gamma_t$ in Fourier space,
we obtain
\begin{eqnarray}
\nonumber
& & \langle\gamma_t^h\rangle(\bm{\theta})\\
\nonumber
&=& -\int
\frac{d^2\mathbf{l}}{(2\pi)^2}
\frac{d^2\mathbf{l}'}{(2\pi)^2}
e^{i\mathbf{l}\cdot\hat{\mathbf n}}e^{i\mathbf{l}'\cdot(\hat{\mathbf
n}+\bm{\theta})}
\cos[2(\phi-\varphi)]
\langle \delta_h({\mathbf l})\kappa({\mathbf
l}')\rangle\\
&=&
-\int
\frac{d^2\mathbf{l}}{(2\pi)^2}
C_l^{h\kappa}
\cos[2(\phi-\varphi)]e^{-i\mathbf{l}\cdot\bm{\theta}},
\end{eqnarray}
where we have used $\langle \delta_h({\mathbf l})\kappa({\mathbf
l}')\rangle=(2\pi)^2C_l^{h\kappa}\delta_D({\mathbf l}+{\mathbf l}')$.
Finally, we take the azimuthal average of $\langle\gamma_t^h\rangle(\bm{\theta})$ to find
the averaged mean tangential shear:
\begin{eqnarray}
\nonumber
 \langle\overline{\gamma}_t^h\rangle(\theta)
&=&\int_0^{2\pi} \frac{d\phi}{2\pi}
\langle\gamma_t^h\rangle(\bm{\theta})\\
\nonumber
&=&
-\int \frac{d^2\mathbf{l}}{(2\pi)^2}
C_l^{h\kappa}
\int_0^{2\pi}\frac{d\phi}{2\pi}
\cos[2(\phi-\varphi)]e^{-il\theta\cos(\phi-\varphi)}\\
\nonumber
&=&
\int \frac{d^2\mathbf{l}}{(2\pi)^2}
C_l^{h\kappa}J_2(l\theta)\\
&=& \int \frac{ldl}{2\pi}
C_l^{h\kappa} J_2(l\theta).
\end{eqnarray}
This completes the derivation of Eq.~(\ref{eq:gammath}).

\section{Derivation of the covariance matrix of the mean tangential
 shear}
\label{app:covariance}

To compute the covariance matrix of the tangential shears (not yet azimuthally
averaged), we first compute
\begin{eqnarray}
\nonumber
& &\langle \gamma_t^h(\bm{\theta})\gamma_t^h(\bm{\theta}')\rangle\\
\nonumber
&=& \frac1{N_L^2}\sum_{ij}^{N_L}\int d^2\hat{\mathbf n}\int
d^2\hat{\mathbf n}'\\
\nonumber
& &\times
\langle
\delta_D(\hat{\mathbf n}-\hat{\mathbf n}_i)
\delta_D(\hat{\mathbf n}'-\hat{\mathbf n}_j)
\gamma_{t}(\hat{\mathbf n}+\bm{\theta})
\gamma_{t}(\hat{\mathbf n}'+\bm{\theta}')
\rangle\\
\nonumber
&=&  \frac1{N_L^2}\int d^2\hat{\mathbf n}\int
d^2\hat{\mathbf n}'\\
\nonumber
& &\times
\left[
\delta_D(\hat{\mathbf n}-\hat{\mathbf n}')
\langle
n_L(\hat{\mathbf n})
\gamma_{t}(\hat{\mathbf n}+\bm{\theta})
\gamma_{t}(\hat{\mathbf n}'+\bm{\theta}')
\rangle\right.\\
& &\left.\quad+
\langle
n_L(\hat{\mathbf n})
n_L(\hat{\mathbf n}')
\gamma_{t}(\hat{\mathbf n}+\bm{\theta})
\gamma_{t}(\hat{\mathbf n}'+\bm{\theta}')
\rangle
\right].
\end{eqnarray}
Here, the first term in the square bracket correlates two $\gamma_t$'s
measured relative to the same lens halo (1-halo term), and the second
correlates two $\gamma_t$'s relative to two lens halos (2-halo term).
Again expanding $n_L$ into the
perturbation, $n_L(\hat{\mathbf n})=\bar{n}_L[1+\delta_h(\hat{\mathbf
n})]$, we obtain
\begin{eqnarray}
\nonumber
& &\langle \gamma_t(\bm{\theta})\gamma_t(\bm{\theta}')\rangle\\
\nonumber
&=&  \frac1{f_{\rm sky}}\frac1{N_L}\int \frac{d^2\hat{\mathbf n}}{4\pi}
 \langle
\gamma_{t}(\hat{\mathbf n}+\bm{\theta})
\gamma_{t}(\hat{\mathbf n}+\bm{\theta}')
\rangle\\
\nonumber
& &+
\frac1{f_{\rm sky}^2}\int \frac{d^2\hat{\mathbf n}}{4\pi}\int
\frac{d^2\hat{\mathbf n}'}{4\pi}
\left[
 \langle
\gamma_{t}(\hat{\mathbf n}+\bm{\theta})
\gamma_{t}(\hat{\mathbf n}'+\bm{\theta}')
\rangle\right.\\
& &
\left.\qquad\quad
+
 \langle
\delta_h(\hat{\mathbf n})
\delta_h(\hat{\mathbf n}')
\gamma_{t}(\hat{\mathbf n}+\bm{\theta})
\gamma_{t}(\hat{\mathbf n}'+\bm{\theta}')
\rangle\right].
\end{eqnarray}
Here, we assume that $\delta_h$ and $\gamma_t$ obey Gaussian statistics, i.e., 
$\langle\delta_h\gamma_t\gamma_t\rangle=0$.
This approximation is justified even in the presence of primordial
non-Gaussianity, as non-Gaussianity is weak, and this approximation
only affects the size of errorbars.
Let us evaluate each term.
With $\gamma_t$ expanded in Fourier space, the first term (1-halo term)
becomes
\begin{eqnarray}
\nonumber
& &
\frac1{N_L}
\frac{1}{f_{\rm sky}}\int\frac{d^2\hat{\mathrm{n}}}{4\pi}
 \langle
\gamma_{t}(\hat{\mathbf n}+\bm{\theta})
\gamma_{t}(\hat{\mathbf n}+\bm{\theta}')
\rangle\\
\nonumber
&=&
 \frac1{N_L}\int \frac{d^2\mathbf{l}}{(2\pi)^2}
C_l^\kappa
\cos\left[2(\phi-\varphi)\right]\cos\left[2(\phi'-\varphi)\right]
e^{i\mathbf{l}\cdot(\bm{\theta}-\bm{\theta}')}\\
& &+\frac{\sigma_\gamma^2}{N_Ln_S}\delta_D(\bm{\theta}-\bm{\theta}'),
\end{eqnarray}
where $\sigma_\gamma$ is the r.m.s. shape noise (dimensionless), and
$n_S$ is the surface density of source (background) galaxies that are
available for the shear measurement at a given location.
By azimuthally averaging $\gamma_{t}$, we find
\begin{eqnarray}
\nonumber
& &
\frac1{N_L}
\int_0^{2\pi}\frac{d\phi}{2\pi}\int_0^{2\pi}\frac{d\phi'}{2\pi}
 \langle
\gamma_{t}(\hat{\mathbf n}+\bm{\theta})
\gamma_{t}(\hat{\mathbf n}+\bm{\theta}')
\rangle\\
\nonumber
&=&
 \frac1{N_L}\int \frac{d^2\mathbf{l}}{(2\pi)^2}
C_l^\kappa J_2(l\theta)J_2(l\theta')\\
& &+\frac{\sigma_\gamma^2}{N_Ln_S}\frac{\delta_D(\theta-\theta')}{2\pi\theta}.
\end{eqnarray}
Here, $C_l^\kappa$ is the angular power spectrum of $\kappa(\mathbf{l})$.

As for the second term (2-halo term),
the first of the second term vanishes, as
$\int d^2\hat{\mathbf n}\gamma_t(\hat{\mathbf n}+\bm{\theta})=0$.
The remaining non-vanishing term gives
\begin{eqnarray}
\nonumber
& & \frac1{f_{\rm sky}^2}\int \frac{d^2\hat{\mathbf n}}{4\pi}\int
\frac{d^2\hat{\mathbf n}'}{4\pi}\\
\nonumber
&\times &
\left[
 \langle
\delta_h(\hat{\mathbf n})
\gamma_{t}(\hat{\mathbf n}+\bm{\theta})
\rangle
\langle
\delta_h(\hat{\mathbf n}')
\gamma_{t}(\hat{\mathbf n}'+\bm{\theta}')
\rangle
\right.\\
\nonumber
& &+
\left.
 \langle
\delta_h(\hat{\mathbf n})
\gamma_{t}(\hat{\mathbf n'}+\bm{\theta}')
\rangle
\langle
\delta_h(\hat{\mathbf n}')
\gamma_{t}(\hat{\mathbf n}+\bm{\theta})
\rangle
\right.\\
\nonumber
& &
\left.
+
 \langle
\delta_h(\hat{\mathbf n})
\delta_h(\hat{\mathbf n}')
\rangle
\langle
\gamma_{t}(\hat{\mathbf n}+\bm{\theta})
\gamma_{t}(\hat{\mathbf n}'+\bm{\theta}')
\rangle
\right]\\
\nonumber
&=&
 \langle
\gamma_t^h(\bm{\theta})
\rangle
\langle
\gamma_t^h(\bm{\theta}')
\rangle\\
\nonumber
& &+
\frac1{4\pi f_{\rm sky}}
\int \frac{d^2\mathbf{l}}{(2\pi)^2}
\cos\left[2(\phi-\varphi)\right]\cos\left[2(\phi'-\varphi)\right]
e^{i\mathbf{l}\cdot(\bm{\theta}-\bm{\theta}')}\\
& &\qquad\qquad\times
\left[(C_l^{h\kappa})^2+C_l^h
\left(C_l^\kappa+\frac{\sigma_\gamma^2}{n_S}\right)\right].
\end{eqnarray}
Here, $C_l^h$ is the angular power spectrum of $\delta_h(\mathbf{l})$.
By azimuthally averaging $\gamma_{t}$ in the above equation, we find
\begin{eqnarray}
\nonumber
& &  \langle
\overline{\gamma}_t^h({\theta})
\rangle
\langle
\overline{\gamma}_t^h({\theta}')
\rangle\\
\nonumber
&+&
\frac1{4\pi f_{\rm sky}}
\int \frac{d^2\mathbf{l}}{(2\pi)^2}J_2(l\theta)J_2(l\theta')\\
& &\qquad\quad\times
\left[(C_l^{h\kappa})^2+C_l^h
\left(C_l^\kappa+\frac{\sigma_\gamma^2}{n_S}\right)\right],
\end{eqnarray}
where we have used the identity
\begin{equation}
\frac{\delta_D(\theta-\theta')}{2\pi\theta}
=\int \frac{ldl}{2\pi}J_2(l\theta)J_2(l\theta').
\end{equation}

Collecting both the 1-halo and 2-halo terms, we finally obtain the
covariance matrix of the azimuthally-averaged mean tangential shear:
\begin{eqnarray}
\nonumber
& &
\langle
 \overline{\gamma}_t^h({\theta})
\overline{\gamma}_t^h({\theta}')
\rangle
-
\langle
 \overline{\gamma}_t^h({\theta})
\rangle
\langle
\overline{\gamma}_t^h({\theta}')
\rangle\\
\nonumber
&=&
\frac1{4\pi f_{\rm sky}}\int \frac{ldl}{2\pi}J_2(l\theta)J_2(l\theta')\\
\nonumber
& &\times
\left[
(C_l^{h\kappa})^2+\left(C_l^h+\frac1{n_L}\right)
\left(C_l^\kappa+\frac{\sigma_\gamma^2}{n_S}\right)
\right].
\end{eqnarray}
This completes the derivation of Eq.~(\ref{eq:cov}).

\section{On the accuracy of Limber's approximation}
\label{app:limber}
Throughout this paper we have repeatedly used Limber's approximation in
order to
relate the angular correlation function to the corresponding three
dimensional power spectrum.
In general, Limber's approximation is known to be accurate only
for small angular scales, and only for the quantities which are
integrated over a broad range of redshift.

However, the situations we have considered
in this paper sometimes violate both of the conditions above:
1) We correlate the convergence field with galaxies within a very thin
redshift slice, and
2) the non-Gaussianity signal we study in this paper appears only on
very large scales.

Then, how accurate is Limber's approximation in this case?
In this Appendix, we shall study in detail the validity and limitation
of Limber's approximation, by comparing the main results of the paper to
the result of \textit{exact} calculations.

Consider a quantity $x_i(\hat{\mathbf{n}})$, which is projected on the
sky. Here,
$\hat{\mathbf{n}}$ is the unit vector pointing toward a given direction
on the sky. This quantity is related to the three dimensional quantity
$s_i(\mathbf{r};z)$
by a projection kernel $W_i(z)$ as
\begin{equation}
x_i(\hat{\mathbf{n}})
=
\int dz W_i(z)s_i[d_A(z)\hat{\mathbf{n}};z].
\label{eq:xi_def}
\end{equation}
Throughout this Appendix, we use $d_A(z)$ to denote $d_A(0;z)$.

Fourier transforming $s_i(\mathbf{r})$, one obtains
\begin{eqnarray}
\nonumber
&&
s_i[d_A(z)\hat{\mathbf{n}};z)]
\\
\nonumber
&=&
\int\frac{d^3k}{(2\pi)^3}
s_i(\mathbf{k},z) e^{i\mathbf{k}\cdot\hat{\mathbf{n}}d_A(z)}
\\
&=&
4\pi\sum_{l,m}i^l
\int\frac{d^3k}{(2\pi)^3}
s_i(\mathbf{k},z)
j_l[kd_A(z)]Y_{lm}^*(\hat{\mathbf{k}})Y_{lm}(\hat{\mathbf{n}}).
\label{eq:si_Fourier}
\end{eqnarray}
In the third line, we have used Rayleigh's formula:
$$
e^{i\mathbf{k}\cdot\hat{\mathbf{n}}r}
=
4\pi\sum_{l,m}i^l
j_l(kr)Y_{lm}^*(\hat{\mathbf{k}})Y_{lm}(\hat{\mathbf{n}}).
$$
By using Eq.~(\ref{eq:si_Fourier}), we rewrite Eq.~(\ref{eq:xi_def}) as
\begin{eqnarray}
\nonumber
x_i(\hat{\mathbf{n}})
&=&
4\pi\sum_{l,m}i^l
\int dz W_i(z)
\\
&\times&
\int\frac{d^3k}{(2\pi)^3}
s_i(\mathbf{k},z)
j_l[kd_A(z)]Y_{lm}^*(\hat{\mathbf{k}})Y_{lm}(\hat{\mathbf{n}}).
\end{eqnarray}
Therefore, the coefficients of the spherical harmonics decomposition of
$x_i(\hat{\mathbf{n}})$, $a_{lm}^{x_i}$, becomes
\begin{equation}
a_{lm}^{x_i}
=
4\pi i^l
\int dz W_i(z)
\int\frac{d^3k}{(2\pi)^3}
s_i(\mathbf{k},z)
j_l[kd_A(z)]Y_{lm}^*(\hat{\mathbf{k}}).
\end{equation}
We calculate the angular power spectrum, $C_l^{x_ix_j}$,
by taking an ensemble average of
$\left\langle a_{lm}^{x_i}a_{lm}^{x_j*} \right\rangle$ as
\begin{eqnarray}
\nonumber
&&
C_l^{x_ix_j}
\\
\nonumber
&\equiv&
\left\langle a_{lm}^{x_i}a_{lm}^{x_j*} \right\rangle
\\
\nonumber
&=&
(4\pi)^2
\int dz W_i(z)
\int dz' W_j(z')
\int\frac{d^3k}{(2\pi)^3}
P^{s_is_j}(\mathbf{k};z,z')
\\
&
&\times
j_l[kd_A(z)]
j_l[kd_A(z')]
Y_{lm}^*(\hat{\mathbf{k}})
Y_{lm}^*(\hat{\mathbf{k}}),
\end{eqnarray}
where we have used the definition of the power spectrum:
$$
\left\langle
s_i(\mathbf{k},z)
s_j^*(\mathbf{k}',z) \right\rangle
\equiv
(2\pi)^3
\delta(\mathbf{k}-\mathbf{k}')
P^{s_is_j}(\mathbf{k};z,z').
$$

Now, by assuming statistical isotropy of the universe, we write
$P^{s_is_j}(\mathbf{k};z,z')=P^{s_is_j}(k;z,z')$, and do the
angular integration of $\hat{\mathbf{k}}$ by using the orthonormality condition
of spherical harmonics:
$$
\int d\hat{\mathbf{k}}
Y_{lm}(\hat{\mathbf{k}})
Y_{lm}^*(\hat{\mathbf{k}}) = 1.
$$
We then obtain the angular power spectrum given by
\begin{eqnarray}
\nonumber
C_l^{x_ix_j}
&=&
\int dz W_i(z)
\int dz' W_j(z')
\\
\nonumber
&&\times
\left\{
\frac{2}{\pi}
\int k^2dk
P^{s_is_j}(k;z,z')
j_l[kd_A(z)]
j_l[kd_A(z')]
\right\}.\\
\label{eq:Cl_exact}
\end{eqnarray}
This is the {\em exact} relation.

What determines the form of $W_i(z)$?
For a projected galaxy distribution projected on the sky, this kernel is
simply a normalized galaxy
distribution function in redshift space. In this paper, we consider the delta
function-like distribution, i.e.,
\begin{equation}
W_g(z)=\delta^D(z-z_L).
\end{equation}
Using Eq.~(\ref{eq:Cl_exact}) with the delta function kernel above yields
Eq.~(\ref{eq:Clgg_exact}):
\begin{equation}
C_l^h
=\frac{2}{\pi}\int dk k^2P_g(k,z_L) j_l^2\left[kd_A(z_L)\right].
\end{equation}
Again, this is still the exact result.
As the form of $W_g(z)$ we have considered here (i.e., a delta function)
is a sharply peaked function, we cannot use Limber's approximation
given below. This is the reason why we have used the exact result for $C_l^h$.

In order to get the expression for Limber's
approximation, we assume that $P^{s_is_j}(k)$
is a slowly-varying function of $k$. Then, by using the identity
\begin{equation}
\frac{2}{\pi}
\int k^2dk
j_l(kr)
j_l(kr')=\frac{\delta^D(r-r')}{r^2},
\end{equation}
we approximate the $k$ integral of Eq.~(\ref{eq:Cl_exact}) as
\begin{eqnarray}
\nonumber
&&
\frac{2}{\pi}
\int k^2dk
P^{s_is_j}(k)
j_l(kr)
j_l(kr')
\\
&\approx&
\frac{\delta^D(r-r')}{r^2}
P^{s_is_j}\left(k=\frac{l+1/2}{r}\right).
\end{eqnarray}
By using this approximation, we finally get
\begin{eqnarray}
\nonumber
C_l^{x_ix_j}
\approx
\int dz W_i(z)W_j(z)
\frac{H(z)}{d_A^2(z)}
P^{s_is_j}\left(k=\frac{l+1/2}{r};z\right),\\
\label{eq:Cl_limber}
\end{eqnarray}
which is the result known as Limber's approximation.

One important application of Limber's approximation is the statistics
involving weak gravitational lensing.
The lensing kernel for the convergence field, $W_\kappa(z)$,
can be calculated by integrating the lens equation:
\begin{equation}
W_\kappa(z)
=\frac{\rho_0}{\Sigma_c(z;z_S)H(z)},
\end{equation}
where $\Sigma_c(z;z_S)$ is the critical surface density defined in
Eq.~(\ref{eq:sigmac}). The exact result for the galaxy-convergence
angular cross power spectrum is
\begin{eqnarray}
\nonumber
C_l^{h\kappa}(z_L)
&=&
\frac{2}{\pi}
\int_0^{z_S} dz
\frac{\rho_0}{\Sigma_c(z;z_S)H(z)}
\\
\nonumber
&\times&
\int dk k^2P_{hm}(k,z_L,z)j_l[kd_A(z_L)]j_l[kd_A(z)],\\
\label{eq:Clhk_exact}
\end{eqnarray}
and the exact result for the convergence-convergence angular
power spectrum is
\begin{eqnarray}
\nonumber
C_l^\kappa(z_S)
&=&\frac{2}{\pi}
\int_0^{z_S}dz
\int_0^{z_S}dz'
\frac{\rho_0^2}{\Sigma_c(z;z_S)H(z)\Sigma_c(z';z_S)H(z')}
\\
&\times&
\int dk k^2P_{m}(k,z,z')j_l[kd_A(z)]j_l[kd_A(z')].
\label{eq:Clkk_exact}
\end{eqnarray}

\begin{figure}
\begin{center}
\includegraphics[width=9cm]{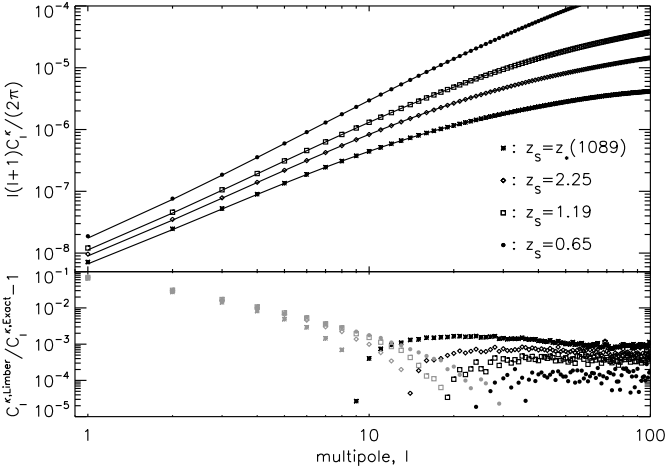}
\end{center}
\caption{
\label{fig:comp_LE_Clkk}
Top: Convergence-convergence angular power spectrum from two different
methods: the exact calculation (Eq.~\ref{eq:Clkk_exact}, symbols)
and Limber's approximation (Eq.~\ref{eq:Clkk_limber}, solid lines).
Bottom: Fractional differences between Limber's approximation and the
exact integration. Symbols are the same as the top panel.
Grey symbols show the absolute values of negative values.
}
\end{figure}
\begin{figure}
\begin{center}
\includegraphics[width=9cm]{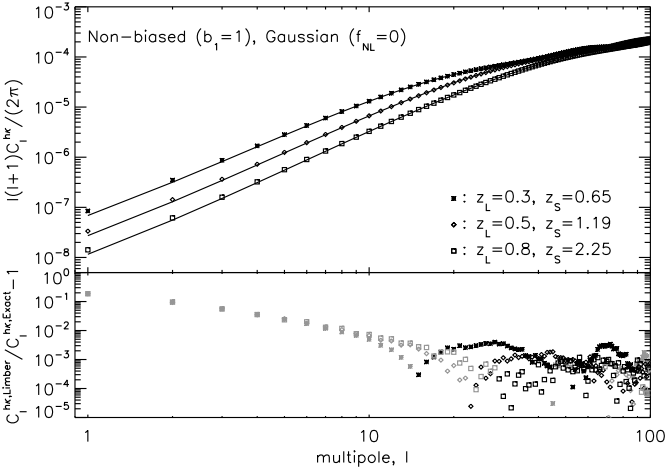}
\end{center}
\caption{
\label{fig:comp_LE_Clhk}
Same as Fig.~\ref{fig:comp_LE_Clkk}, but for the
galaxy-convergence cross angular power spectrum
with $\fnl=0$ and $b_1=1$.
}
\end{figure}
\begin{figure}
\begin{center}
\includegraphics[width=9cm]{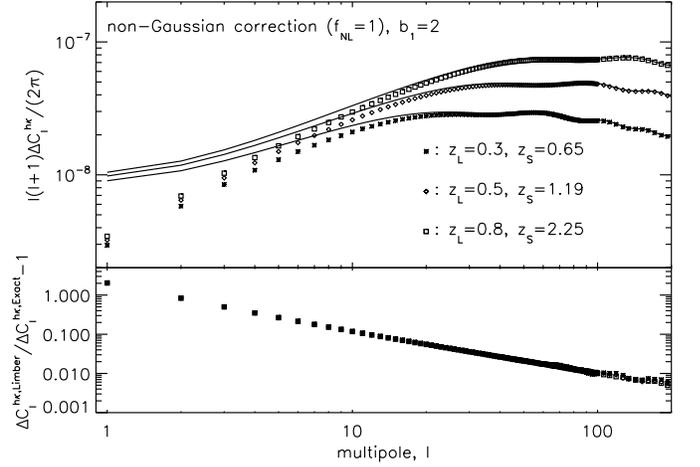}
\end{center}
\caption{
\label{fig:comp_LE_Clhk_nG}
Same as Fig.~\ref{fig:comp_LE_Clkk}, but for the
non-Gaussian correction (i.e., the term proportional to $\Delta b(k)$)
 to the galaxy-convergence cross angular power spectrum.
We show the corrections with $\fnl=1$ and $b_1=2$.
}
\end{figure}
\begin{figure}
\begin{center}
\includegraphics[width=9cm]{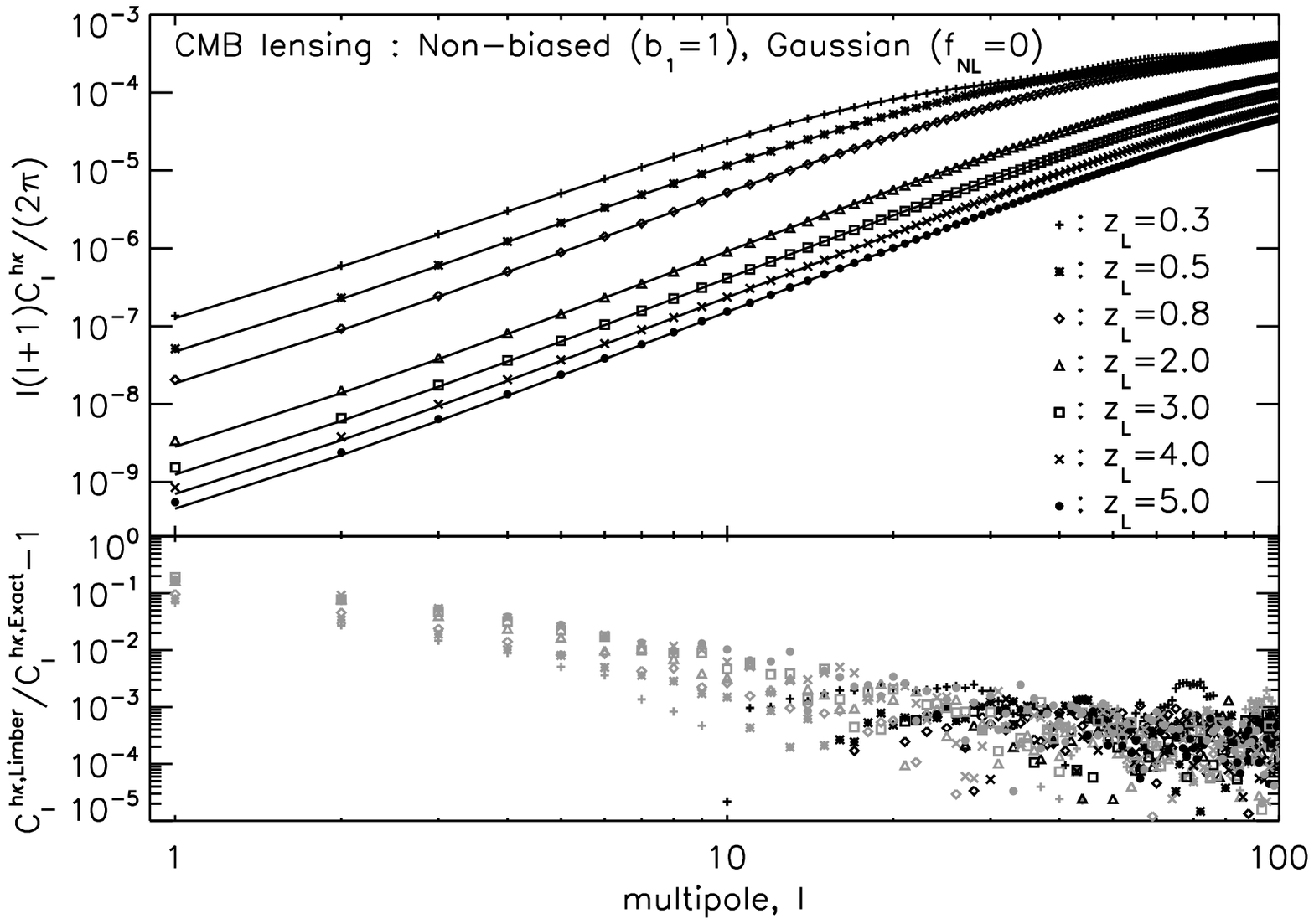}
\end{center}
\caption{
\label{fig:comp_LE_cmblens}
Same as Fig.~\ref{fig:comp_LE_Clhk}, but for the galaxy-CMB lensing.
}
\end{figure}
\begin{figure}
\begin{center}
\includegraphics[width=9cm]{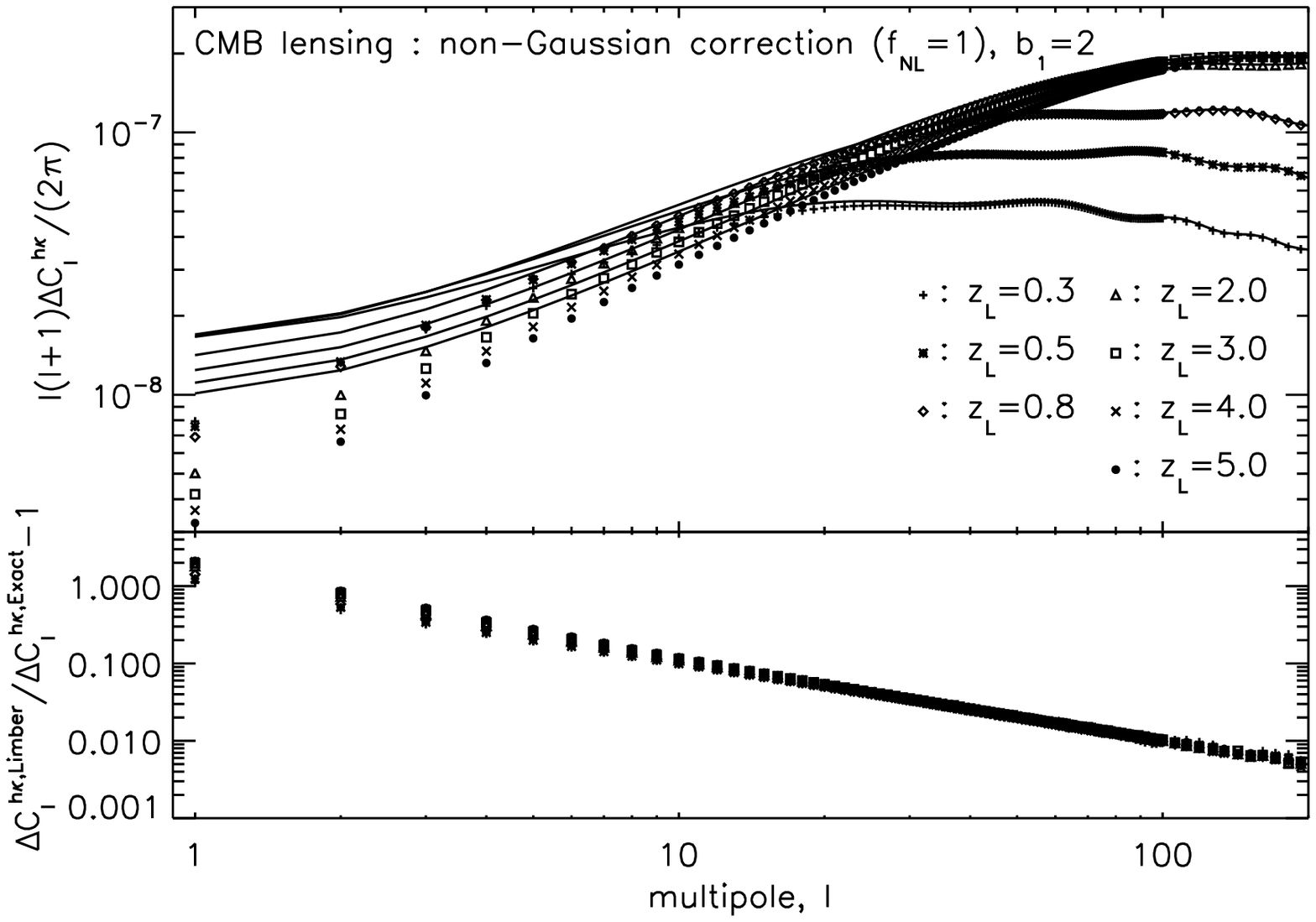}
\end{center}
\caption{
\label{fig:comp_LE_cmblens_nG}
Same as Fig.~\ref{fig:comp_LE_Clhk_nG}, but for the galaxy-CMB lensing.
}
\end{figure}
\begin{figure}
\begin{center}
\includegraphics[width=9cm]{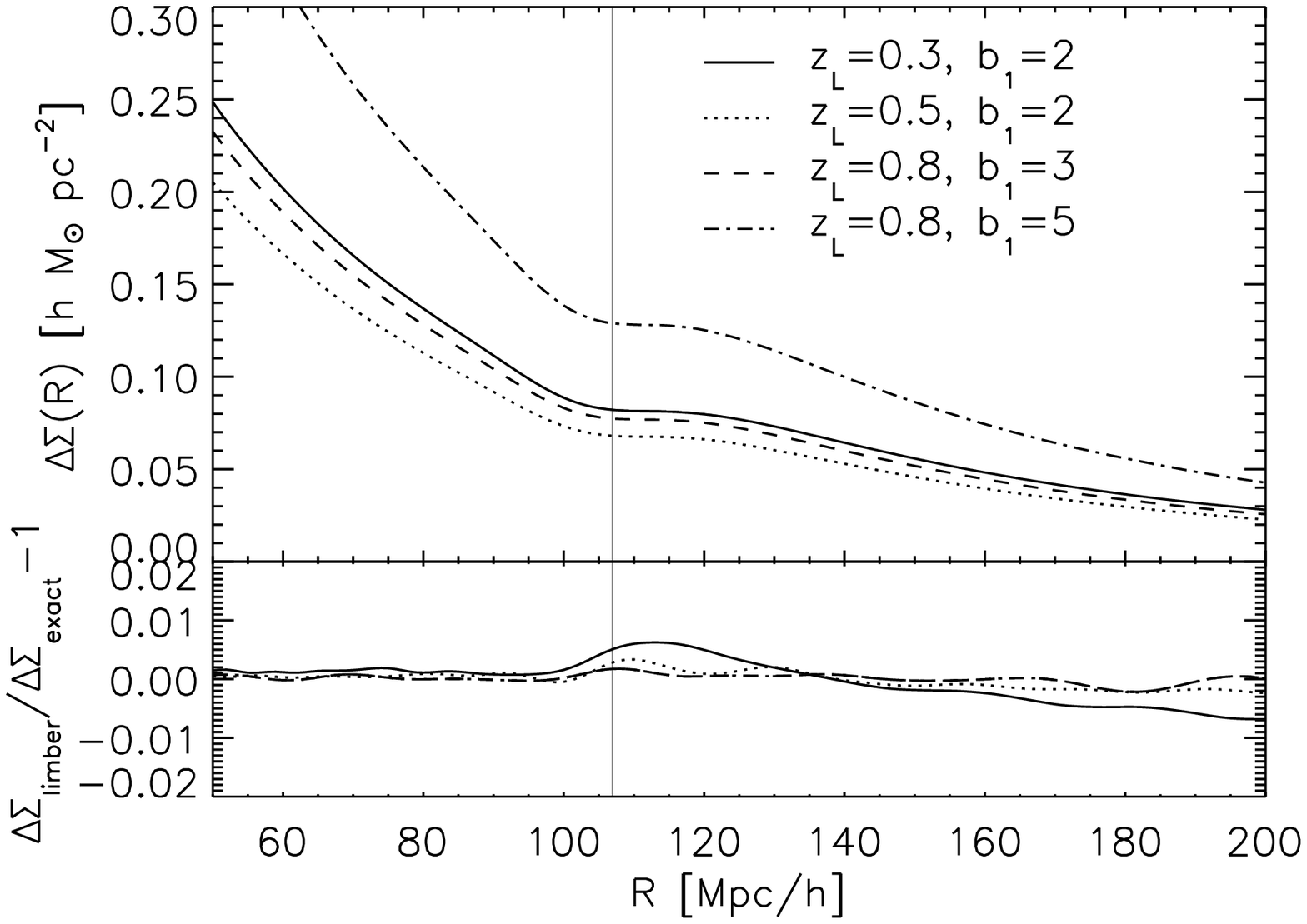}
\end{center}
\caption{
\label{fig:comp_LE_bao}
Top: Same as Fig.~\ref{fig:bao}, but also showing the exact result
 (Eq.~\ref{eq:Clhk_exact}, thick lines) on top of the result from
Limber's approximation (Eq.~\ref{eq:Clhk_limber}, thin lines).
Bottom: Fractional difference of Limber's approximation relative to the
exact result.
}
\end{figure}

First, we compare the exact convergence-convergence angular power
spectrum to Limber's approximation.
Fig.~\ref{fig:comp_LE_Clkk}
shows that  Limber's approximation works very well
for all four source redshifts we study in the paper: $z_S=0.65$,
$1.19$, $2.25$, and $1089.0$.
For $l>10$, the error caused by Limber's approximation is always much
smaller than 1\%.

Then, we compare the galaxy-convergence cross angular power spectra.
Fig.~\ref{fig:comp_LE_Clhk} and Fig.~\ref{fig:comp_LE_Clhk_nG}
show the comparison between the exact galaxy-convergence
cross power spectrum (Eq.~\ref{eq:Clhk_exact}, symbols) and
their Limber approximation (Eq.~\ref{eq:Clhk_limber}, solid lines) for
three galaxy-galaxy lensing cases we study in Sec.~\ref{sec:gglens}:
$(z_L,z_S)=(0.3,0.65)$, $(0.5,1.19)$, and $(0.8,2.25)$.

For the Gaussian term (Fig.~\ref{fig:comp_LE_Clhk}), Limber's
approximation is accurate at $l>10$ with the errors less than 1\%.
On the other hand, Limber's approximation to the non-Gaussian correction term
(Fig.~\ref{fig:comp_LE_Clhk_nG}) has a sizable error, at the level of
10\%, at $l\sim 10$. The error goes down to the 1\% level only at $l\sim
100$. One needs to keep this in mind when comparing Limber's
approximation with observations.
We find that Limber's approximation underpredicts
the Gaussian term at $l\lesssim 20$, while it overpredicts  the
non-Gaussian corrections at all multipoles.

The story is basically the same for the galaxy-CMB lensing cross power spectrum.
Fig.~\ref{fig:comp_LE_cmblens} (Gaussian term) and
Fig.~\ref{fig:comp_LE_cmblens_nG} (non-Gaussian correction)
show the comparison between
the exact galaxy-convergence cross power spectrum
(Eq.~\ref{eq:Clhk_exact}, solid lines) and
their Limber approximation (Eq.~\ref{eq:Clhk_limber}, dashed lines) for
seven lens redshifts we study in Sec.~\ref{sec:cmblens}:
$z_L=0.3$, $0.5$, $0.8$, $2$, $3$, $4$, and $5$.
Again, for small scales, $l>10$, Limber's approximation works better than
1\% for the Gaussian term, while it overpredicts the non-Gaussian
correction at the level of 10\% at $l\sim 10$ and 1\% at $l\sim 100$.

\begin{figure}
\begin{center}
\includegraphics[width=9cm]{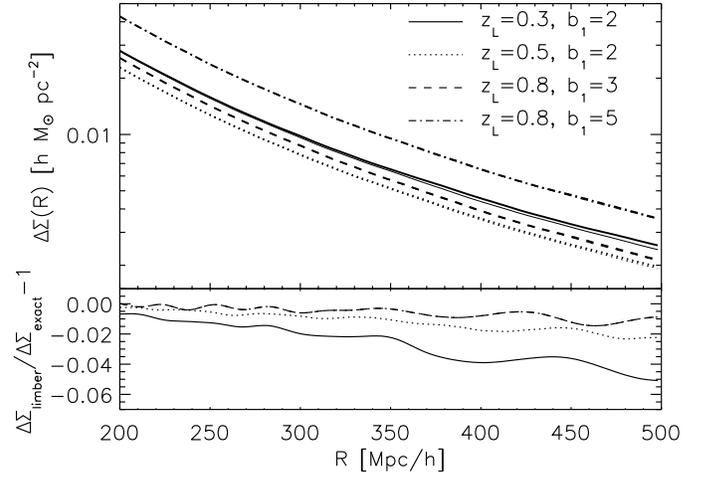}
\end{center}
\caption{
\label{fig:comp_LE_DS_highR}
Same as Fig.~\ref{fig:comp_LE_bao}, but for larger $R$.
Thick lines are the results of the exact integration, while the thin
lines are Limber's approximation. The Limber approximation overpredicts
$\Delta\Sigma(R)$ for large $R$, but the error is at most 5\% for
 $R<500~h^{-1}~{\rm Mpc}$.
The error is the largest for the lowest $z_L$, as a  physical
 separation $R$ at a lower  redshift
corresponds to a larger angular separation on the sky.
}
\end{figure}
\begin{figure}
\begin{center}
\includegraphics[width=9cm]{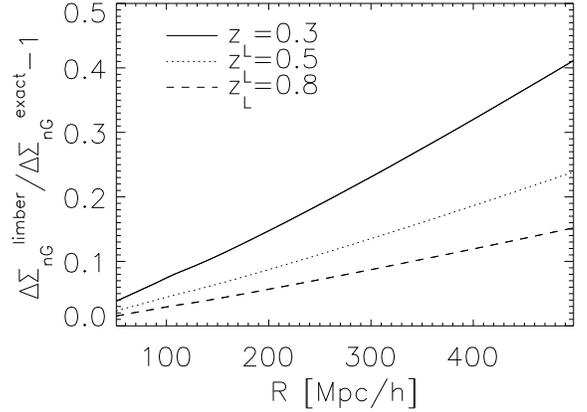}
\end{center}
\caption{
\label{fig:comp_LE_DS_nG_comp}
Fractional differences in the non-Gaussian correction terms,
$\Delta\Sigma_\mathrm{nG}$, from Limber's approximation and the exact
integration. Using Limber's approximation, we overpredict the non-Gaussian
correction by $\sim 20\%$ at $R=300~h^{-1}~\mathrm{Mpc}$ for $z_L=0.3$.
}
\end{figure}

What about the effect on the mean tangential shear, $\Delta\Sigma(R)$?
Fig.~\ref{fig:comp_LE_bao} compares the Gaussian term of
$\Delta\Sigma(R)$ from
the exact integration and that from Limber's approximation.
On the top panel of
Fig.~\ref{fig:comp_LE_bao}, we show the baryonic feature computed with
Limber's approximation (thin lines, the same as those in
Fig.~\ref{fig:bao}) as well as that computed with the exact integration
(thick lines). They are indistinguishable by eyes. The bottom panel
shows
 the fractional differences between the two. We find that
Limber's approximation is better than $0.5\%$ for
$R<180~h^{-1}~\mathrm{Mpc}$; thus, the baryonic feature in
$\Delta\Sigma$ is not an artifact caused by Limber's approximation.

However, Limber's approximation becomes worse and worse as we go to
larger $R$. Fig.~\ref{fig:comp_LE_DS_nG_comp} shows $\Delta\Sigma$ on
large scales. For the lens redshifts that we have studied here, the
error is at most 5\% for  $R<500~h^{-1}~{\rm Mpc}$, and the error is the
largest for the lowest $z_L$, as a given $R$ at a lower redshift
corresponds to a larger angular separation on the sky.

While Limber's approximation underpredicts the Gaussian term on large
scales, it overpredicts the non-Gaussian correction
terms. Fig.~\ref{fig:comp_LE_DS_nG_comp} shows the fractional
differences of the non-Gaussian correction terms,
$\Delta\Sigma_\mathrm{nG}$, between Limber's approximation and the exact
calculation as a function of separation $R$  for three lens redshifts:
$z_L=0.3$, $0.5$, and $0.8$. This figure shows that the error caused by
Limber's approximation can be substantial on $\Delta\Sigma_\mathrm{nG}$.

\begin{figure}
\begin{center}
\includegraphics[width=8cm]{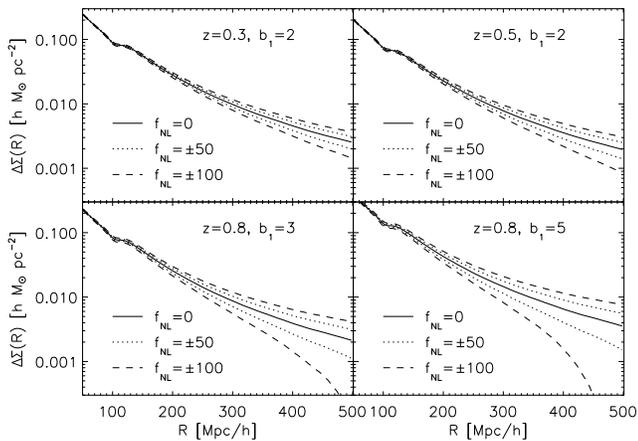}
\end{center}
\caption{
\label{fig:comp_LE_DS_nG}
Same as Fig.~\ref{fig:fnl_gglens_signal}, but with the exact integration
instead of Limber's approximation.
}
\end{figure}

As Limber's approximation to $\Delta\Sigma(R)$ can be quite inaccurate
on very large scales,  we show the exact calculations of
$\Delta\Sigma(R)$ in Fig.~\ref{fig:comp_LE_DS_nG}.
(Limber's approximation is given in Fig.~\ref{fig:fnl_gglens_signal}.)

Finally, we note that the definition of the tangential shear we have
used (Eq.~\ref{eq:gammat}) is valid only on the flat sky (as noted in
the footnote there), and thus the prediction for $\Delta\Sigma$ on very
large scales probably needs to be revisited with the exact definition of the
tangential shears on the full sky using the spin-2 harmonics. This is
beyond of the scope of our paper.

\end{document}